\documentclass{ar}

\usepackage[numbers]{natbib}
\usepackage{amsmath,amssymb,amsbsy}
\usepackage{abraces}
\usepackage{isotope}
\usepackage{physics}
\usepackage{dsfont}
\usepackage[capitalize,noabbrev]{cleveref}
\usepackage{url}
\usepackage{graphicx}

\definecolor{joelred}{rgb}{0.81,0.13,0.16}

\newcommand{\tdott}[2]{\vb*{\tau}_{#1}\vdot\vb*{\tau}_{#2}}
\newcommand{\sdots}[2]{\vb*{\sigma}_{#1}\vdot\vb*{\sigma}_{#2}}
\newcommand{\eu}[1]{\mathrm{e}^{#1}}
\newcommand{\ii}{\mathrm{i}}

\crefformat{equation}{Equation~#2#1#3}
\crefformat{figure}{\textbf{Figure~#2#1#3}}
\crefformat{table}{Table~#2#1#3}
\crefformat{section}{Section~#2#1#3}

\jname{Annu. Rev. Nucl. Part. Sci.}
\jvol{AA}
\jyear{2019}
\doi{10.1146/((please add article doi))}

\begin{document}

\markboth{Lynn et al.}{Quantum Monte Carlo Methods in Nuclear Physics: Recent Advances}

\title{Quantum Monte Carlo Methods in Nuclear Physics: Recent Advances}

\author{J.~E.~Lynn,$^{1,2}$ I.~Tews,$^3$ S.~Gandolfi,$^3$ and A.~Lovato$^{4,5}$
\affil{$^1$Institut f\"ur Kernphysik,
Technische Universit\"at Darmstadt, 64289 Darmstadt, Germany; email: joel.lynn@physik.tu-darmstadt.de}
\affil{$^2$ExtreMe Matter Institute EMMI, GSI Helmholtzzentrum f\"ur
Schwerionenforschung GmbH, 64291 Darmstadt, Germany}
\affil{$^3$Theoretical Division, Los Alamos National Laboratory, Los Alamos, NM 87545, USA; email: itews@lanl.gov, stefano@lanl.gov}
\affil{$^4$Physics Division, Argonne National Laboratory, Argonne, Illinois 60439, USA; email: lovato@anl.gov}
\affil{$^5$INFN-TIFPA Trento Institute of Fundamental Physics and Applications, Via Sommarive, 14, 38123 Trento, Italy}
}

\begin{abstract}
In recent years, the combination of precise quantum Monte Carlo (QMC) methods with realistic nuclear interactions and consistent electroweak currents, in particular those constructed within effective field theories (EFTs), has lead to new insights in light and medium-mass nuclei, neutron matter, and electroweak reactions. This compelling new body of work has been made possible both by advances in QMC methods for nuclear physics, which push the bounds of applicability to heavier nuclei and to asymmetric nuclear matter and by the development of local chiral EFT interactions up to next-to-next-to-leading order and minimally nonlocal interactions including $\Delta$ degrees of freedom. In this review, we discuss these recent developments and give an overview of the exciting results for nuclei, neutron matter and neutron stars, and electroweak reactions.
\end{abstract}

\begin{keywords}
many-body methods, nuclear interactions, chiral effective field theory, quantum Monte Carlo methods, light and medium-mass nuclei, electroweak properties of nuclei
\end{keywords}
\maketitle

\tableofcontents

\section{INTRODUCTION}
Over the past decade, significant progress has been made in
the theoretical description of strongly-interacting nuclear
systems. This progress is reflected in an
increasingly accurate prediction of nuclear-structure observables
for heavier nuclei, including, e.g. radii, masses, and
neutron-separation energies.
In particular, compelling progress has been 
made in \textit{ab initio} nuclear structure, where the many-body Schr\"odinger equation is solved with controlled approximations and protons and neutrons are assumed to be
the relevant degrees of freedom.
This progress includes, for example, the discovery of new shell 
closures in neutron-rich nuclei, new studies of doubly magic nuclei,
and the role of short-range correlations in weak
transitions~\cite{Simonis:2017dny,Hergert:2015awm,Leistenschneider:2017mrr,
Hagen:2016uwj,Morris:2017vxi,Pastore:2017uwc}.
In addition, the description of nuclear and neutron matter, and therefore the
symmetry energy, has also become more accurate, and now includes reliable
uncertainty estimates that are important for the extrapolation to the
density regime encountered in neutron stars~\cite{Gandolfi:2012,Tews:2018kmu}. 
These advances have
been steered by the developments of systematic nuclear interactions and
reliable many-body methods.

Systematic Hamiltonians from nuclear effective field theories
(EFTs)~\cite{Bedaque:2002mn,Epelbaum:2008ga,Machleidt:2011zz} have
played a key role in obtaining reliable results for nuclear systems.
These EFT Hamiltonians are rooted in the symmetries of the fundamental
theory of strong interactions, quantum chromodynamics (QCD), but
describe the dynamics of nuclear system in terms of nucleonic degrees of
freedom.\begin{marginnote}[]\entry{EFT}{effective field theory}
\entry{QCD}{quantum chromodynamics} \end{marginnote}They are based on a
power-counting scheme that allows for the derivation of nuclear
interactions and consistent electroweak currents in a systematically
improvable fashion. Furthermore, nuclear EFTs provide a recipe to
estimate theoretical uncertainties, a key ingredient for a meaningful
comparison with experimental data. In addition, nuclear EFTs naturally
predict many-body forces, which are necessary for the correct
description of nuclear systems. Pionless EFT and chiral (pion-full) EFT,
both used in QMC calculations, will be briefly discussed in this review.

Thanks to the increasing availability of computing resources and the
development of new algorithms, nuclear  \textit{ab initio} methods have
extended their reach to medium-heavy nuclei. Among these many-body
methods, quantum Monte Carlo (QMC) techniques are known for their
accuracy in describing properties of light nuclei up to \isotope[12]{C},
see e.g. References~\cite{Carlson:2014vla,Epelbaum:2011md}.
Among these QMC approaches are the Green's Function Monte Carlo (GFMC) and the auxiliary field diffusion Monte Carlo (AFDMC) methods.
They solve the Schr\"odinger equation by exploiting an imaginary-time
evolution to enhance the ground-state component from a starting trial
wave function. While some approximations are made during this evolution,
mainly to cope with the fermion sign problem, the final results can be
considered ``stochastically exact'', as expectation values are estimated
on finite Monte Carlo
samples.\begin{marginnote}[]\entry{QMC}{quantum Monte Carlo}\entry{GFMC}{Green's function Monte Carlo}\entry{AFDMC}{auxiliary field diffusion Monte Carlo}\end{marginnote}The
GFMC method has been used to successfully predict the spectra and electroweak
processes of nuclei with $A\leq 12$, where $A$ is the number of nucleons,
with a percent-level accuracy. Because it sums over all spin/isospin
states, the GFMC scales exponentially with $A$, which
presently prevents its applicability to $A>12$ nuclei. The AFDMC method,
on the other hand, uses Hubbard-Stratonovich transformations to
sample the spin/isospin degrees of freedom and achieve a polynomial
scaling in $A$. This has enabled the computation of systems with larger
$A$, such as \isotope[16]{O} and neutron matter, at the cost of using
somewhat simplified wave functions.

In this review, we present state-of-the-art QMC results
for nuclei up to \isotope[16]{O} and neutron matter using
nuclear EFT interactions, and electroweak processes using realistic
phenomenological potentials, and discuss future directions.
The review is structured as follows. In~\cref{sec:interactions} we
discuss nuclear interactions: Starting with phenomenological ones for
context and moving to those based on chiral EFT.
In~\cref{sec:currents} we briefly discuss electroweak
currents. In~\cref{sec:QMC} we introduce the GFMC and AFDMC methods
for nuclear physics. In~\cref{sec:nuclei,sec:matter,sec:electroweak}
we present recent results for light and medium-mass nuclei, neutron
matter and neutron stars, as well as for electroweak reactions.

\section{NUCLEAR INTERACTIONS}\label{sec:interactions}

The fundamental degrees of freedom for nuclear systems are quarks
and gluons, whose dynamics are primarily governed by the QCD
Lagrangian. However, a description of, e.g., atomic nuclei, in
terms of these degrees of freedom requires nonperturbative lattice
techniques. Because of their tremendous computing cost, they are
currently not practical for $A\gtrsim2$ at a physical value of the pion
mass~\cite{Savage:2011xk}.

Instead, at the energy regime relevant for the description of nuclear systems, the effective degrees of freedom are point-like nucleons, whose dynamics are dictated by the nonrelativistic Hamiltonian
\begin{equation}\label{eq:Hamiltonian}
H=T+\sum_{i<j} V_{ij}^\text{NN}+\sum_{i<j<k} V_{ijk}^\text{3N}\,+\cdots .
\end{equation} 
In the above equation, $T$ denotes the kinetic energy, $V_{ij}^\text{NN}$ is the two-nucleon (NN) interaction between nucleons $i$ and $j$, $V_{ijk}^\text{3N}$ is the three-nucleon (3N) interaction between nucleons $i$, $j$, and $k$, and the ellipsis indicate interactions involving more than three particles. As suggested by nuclear matter studies~\cite{Kruger:2013kua}, these four-nucleon (and beyond) interactions are small compared to the current level of precision and can safely be
omitted.\begin{marginnote}[]\entry{NN}{two-nucleon}\entry{3N}{three-nucleon}\end{marginnote}

Traditionally, nuclear interactions have been constructed
relying on meson-exchange models, e.g. in the CD-Bonn
potential~\cite{Machleidt:2000ge}, or with the goal of reproducing
scattering data with very high accuracy, e.g., nuclear interactions of the
Argonne type~\cite{Wiringa:1994wb}. Phenomenological Argonne NN interactions
have been extensively and successfully used in a number
of GFMC and AFDMC calculations. They describe the NN interaction by
explicitly including the long-range one-pion exchange (OPE)
interaction and a set of intermediate- and short-range terms that model
the more complicated multi-pion exchanges and short-range dynamics. The
OPE is given by\begin{marginnote}[]\entry{OPE}{one-pion exchange}\end{marginnote}
\begin{align}
V_{ij}^{\text{NN},\pi}=\frac{f_\pi^2}{4\pi}\frac{m_{\pi}}{3}\left[Y(m_{\pi}r) \sdots{i}{j}+T(m_{\pi}r)Y(m_{\pi}r) S_{ij} \right]\tdott{i}{j}\,,
\end{align}
where $f_\pi$ is the $\pi\text{N}$ coupling constant,
$m_{\pi}$ is the average pion mass, $S_{ij}=3 \vb*{\sigma}_i\vdot
\vb{\hat{r}}\,\vb*{\sigma}_j\vdot\vb{\hat{r}}-\sdots{i}{j}$ is the tensor
operator in coordinate space with the Pauli matrices $\vb*{\sigma}$, $Y(x)=\exp(-x)/x$ is the Yukawa function
and $T(x)=(1+3/x+3/x^2)$. The short-range divergent behavior of both $T$
and $Y$ is regulated by multiplying them by $f_R(x)=1-\exp(-cx^2)$ where
$c$ is a cutoff parameter, typically taken to be $c=2.1$~$\text{fm}^{-2}$.
The intermediate- and short-range parts are modeled by a set of
spin-/isospin- and momentum-dependent operators multiplied by $T^2$
(to approximate the two-pion-exchange) and Woods-Saxon-like radial
functions respectively. The latest version in this class of potentials,
denoted as Argonne $v_{18}$ (AV18)~\cite{Wiringa:1994wb}, is expressed
in terms of 18 operators:
\begin{align}
O_{ij}^{1-8}&=\left\lbrace \mathds{1},\sdots{i}{j},S_{ij}, \vb{L}\vdot \vb{S}\right\rbrace\times \left\lbrace\mathds{1},\tdott{i}{j} \right\rbrace\,,\\
O_{ij}^{9-14}&=\left\lbrace\vb{L}^2, \vb{L}^2\sdots{i}{j}, (\vb{L}\vdot \vb{S})^2\right\rbrace \times\left\lbrace\mathds{1},\tdott{i}{j}\right\rbrace\,,\\
O_{ij}^{15-18}&=\left\lbrace T_{ij}, \sdots{i}{j}T_{ij},S_{ij}T_{ij}, \tau_i^z+\tau_j^z\right\rbrace\,,\label{eq:operators}
\end{align}
In the above equations, $\vb{L}$ is the relative angular
momentum of the pair, $\vb{S}$ is the total spin, and
$T_{ij}=3\tau_i^z\tau_j^z-\tdott{i}{j}$ is the isotensor operator. All
parameters of AV18 have been fit to the Nijmegen NN scattering database
with $\chi^2/{\rm datum}\simeq 1$. 
Simplified versions of these interactions, comprising only a subset of
the operators reported in~\cref{eq:operators} are available. For
instance, the Argonne $v_8'$ ($\text{AV8}'$) interaction, widely used in
neutron-matter studies, only contains
the first 8 operators, and other even simpler interactions have
been explored~\cite{Wiringa:2002ja}. In addition to the NN forces, phenomenological 3N interactions have been developed. They are generally expressed as a sum of a two-pion-exchange $P$-wave term (Fujita-Miyazawa), a two-pion-exchange $S$-wave contribution, a three-pion-exchange contribution, and a 3N contact.
More specifically, the Urbana IX (UIX)~\cite{Carlson:1983kq} interaction
contains only the first and last terms, while the Illinois 7
(IL7)~\cite{Pieper:2001ap} potential contains all four contributions.
The UIX 3N interaction is fit to reproduce the ground-state energies of
\isotope[3]{H} and \isotope[4]{He} and the saturation-point of symmetric
nuclear matter, while the IL7 interaction was fit to the low-lying
spectra of nuclei in the mass range $A=\text{3--10}$. Phenomenological
interactions have been successfully used in a multitude of QMC
calculations of nuclear systems, and results have been reviewed in e.g.,
Reference~\cite{Carlson:2014vla}. 

However, these interactions suffer from
important shortcomings. Since they are constructed in an empirical
way without a clear guiding principle, it is not possible to assess
theoretical uncertainties associated with modeling nuclear dynamics. Also,
it is not clear how to improve these interactions, especially in the
3N sector. For example, though the AV18+IL7 Hamiltonian leads to a
description of more than 100 ground- and excited-state energies up to
$A=12$ in good
agreement with experimental data, it fails to provide
sufficient repulsion in pure neutron
matter~\cite{Maris:2013rgq}. On the other hand, the AV18+UIX model,
while providing a reasonable description of nuclear matter properties,
does not satisfactorily reproduce the spectrum of light nuclei. In
addition, the derivation of consistent electroweak currents is not
straightforward.

A solution to the previously discussed limitations of phenomenological
interactions has been presented with the advent of nuclear
EFTs~\cite{Bedaque:2002mn,Epelbaum:2008ga,Machleidt:2011zz}. Nuclear
EFTs exploit the symmetries of QCD and enable a systematic approach
to nuclear forces based on a low-momentum expansion. Within the nuclear
EFT approach, one starts from hadronic degrees of freedom relevant for
the system at hand. Additional degrees of freedom, e.g. heavier mesons
or even the nucleon
substructure, relevant only at higher energy scales beyond those treated
within the EFT, are integrated out. This so-called ``separation of
scales'' determines the breakdown scale, $\Lambda_b$, of the theory,
and can be used to construct a systematic EFT: One writes down
the most general Lagrangian consistent with all symmetries of QCD, and
uses a power-counting scheme to arrange the terms according to their
importance, typically in powers of $p/\Lambda_b$, where $p$ is a typical
momentum scale in the nuclear system. The resulting scheme
is valid only when $p\ll \Lambda_b$.

The most general EFT Lagrangian contains an infinite series of interaction terms,
\begin{align}
V=\sum_{\nu=0}^{\infty}V^{\nu}(C_i^{\nu}) \left(\frac{p}{\Lambda_b} \right)^{\nu}\,, \label{eq:EFTSeries}
\end{align}
where $V^{\nu}(C_i^{\nu})$ is the contribution at order $\nu$ which
depends on low-energy couplings (LECs) $C_i^{\nu}$. The LECs encode the unresolved physics that is integrated out and are determined by
fitting experimental data.\begin{marginnote}[]\entry{LECs}{low-energy couplings}\end{marginnote}In
a converging EFT, the LECs are natural, i.e. of order $1$ and, hence, the higher-order contributions to $V$ decrease in
magnitude. This permits the truncation of the series expansion at a certain order $\nu$. By going to higher orders, one can
work to a desired accuracy at the cost of computing more diagrams. 
(In this review we denote leading order by LO, next-to-leading order by NLO, and $\text{next-to-}\cdots\text{-leading order}$ by N$^x$LO, with $x$ the number of orders beyond LO).
\begin{marginnote}[]\entry{LO}{leading order}\entry{NLO}{next-to-leading order}\entry{N$^{\boldsymbol{x}}$LO}{$\overbrace{\text{next-to-}\cdots}^{x\
\text{times}}$~-leading order}
\end{marginnote}This systematic expansion can be used to estimate meaningful theoretical uncertainties. Another advantage of nuclear EFTs is that the procedure described above leads to the natural appearance of many-body forces.

At very low momentum scales, $p\ll m_{\pi}$, pions can be integrated out
and nuclear interactions reduce to contact interactions with different
numbers of derivatives.
QMC calculations with
pionless EFT interaction have been used to analyze lattice QCD
calculations with great success~\cite{Contessi:2017rww,Barnea:2013uqa}, e.g., a pionless-EFT Hamiltonian was used in AFDMC
calculations aimed at extending lattice QCD predictions to
\isotope[16]{O}~\cite{Contessi:2017rww}. Leading-order results indicate
that for $m_{\pi}=805$~MeV and $m_{\pi}=510$~MeV, \isotope[16]{O} is not
stable against breakup into four \isotope[4]{He} nuclei. We refer the
reader to References~\cite{Bedaque:2002mn,Epelbaum:2008ga} for more
details on pionless EFT.  

While pionless EFT has been used successfully in low-energy nuclear
physics, see, e.g., Reference~\cite{Bansal:2017pwn}, typical momenta in
nuclear many-body systems are of the order of $m_{\pi}$, and therefore
larger than its breakdown scale. 
Chiral EFT is based on the observation that pions naturally emerge as pseudo
Goldstone bosons associated with the spontaneous breaking of the approximate
chiral symmetry of QCD. Within this ``pionfull'' chiral EFT~\cite{Epelbaum:2008ga,Machleidt:2011zz},
nuclear interactions are comprised of both contact terms, written in a
general operator basis, and one- and multi-pion-exchange interactions.
Modern chiral EFT interactions are based on Weinberg power
counting~\cite{Weinberg:1990rz,Weinberg:1991um}, but alternative
power-counting schemes have been suggested; see, e.g.,
References~\cite{Kaplan:1998tg,Kaplan:1998we,Nogga:2005hy,PavonValderrama:2005wv,Long:2011xw}.

At LO, the contact interactions are given by the momentum-independent contributions
\begin{align}
V_{\text{cont}}^{\nu=0}=C_{\mathds{1}}\mathds{1}+C_{\sigma}\sdots{1}{2}+C_{\tau}\tdott{1}{2}+C_{\sigma\tau}\sdots{1}{2}\tdott{1}{2}\,,
\end{align}
and the pion-exchange interactions are given by the well-known OPE interaction,
\begin{align}
V_{\pi}^{\nu=0}(\textbf{p}, \textbf{p}')=-\left(\frac{g_A}{2 f_\pi} \right)^2 \frac{\vb*{\sigma}_1\vdot\vb{q}\,\vb*{\sigma}_2\vdot\vb{q}}{\vb{q}^2+m_{\pi}^2}\tdott{1}{2}\,,
\end{align}
where $\vb{p}$ and $\vb{p}'$ are the relative nucleon momenta 
before and after the
interaction, and $\vb{q}=\vb{p}-\vb{p}'$ is the momentum transfer.
At higher orders, more complicated interaction pieces contribute, i.e.,
momentum-dependent contacts $\sim p^{\nu}, p'^{\nu}$, tensor
contacts, and multiple-pion exchanges, and we refer the reader to References~\cite{Epelbaum:2008ga,Machleidt:2011zz} for more details.
For example, at N$^3$LO the NN operator basis includes a set of operators
similar to the phenomenological ones of~\cref{eq:operators}, but the
operators in the contact and pion sector appear in a systematic fashion.
A chief advantage of the chiral EFT formulation is that 3N interactions
are consistent with the NN potential, i.e., the same vertices in both
sectors have the same LECs. The leading 3N forces are given by a
two-pion exchange interaction ($V_C$), an OPE--contact interaction ($V_D$) and a 3N contact contribution ($V_E$), with only two unknown LECs to be determined~\cite{vanKolck:1994yi}.

Chiral EFT Hamiltonians have been extensively employed in recent years
in a host of nuclear many-body methods. At the same time, new strategies
to improve chiral interactions and reduce the theoretical uncertainties
have been proposed.  Interactions up to fifth order in the chiral
expansion have recently been developed~\cite{Reinert:2017usi,Entem:2017gor}. These interactions
reproduce the pp and np scattering data from the Granada-2013 database
with a $\chi^2/\text{datum}\sim1$, matching the precision of
phenomenological potentials. Furthermore, potentials with explicit
$\Delta$ degrees of freedom have been
constructed~\cite{Piarulli:2014bda,Ekstrom:2017koy} which in principle
enable a detailed comparison of the order-by-order convergence in both
the $\Delta$-less and $\Delta$-full theories.  New optimization schemes
are being explored~\cite{Ekstrom:2015rta,Carlsson:2015vda},
which may improve the multidimensional fits of the LECs in the chiral
Hamiltonians. Such schemes may play an important role when working at N$^3$LO or
beyond, where 24 or more LECs have to be simultaneously determined.
Also, new forms of uncertainty estimates using Bayesian statistical
tools are being explored~\cite{Furnstahl:2015rha,Melendez:2017phj}.
These tools are necessary to enable a meaningful comparison of
theoretical calculations with experimental data and to study the
convergence of the chiral expansion in a systematic way.  Lastly, new
regularization schemes are being explored. As for phenomenological
interactions, when employing nuclear EFT Hamiltonians in many-body
calculations, the high-momentum components of the interactions have to
be regularized to prevent divergences. This is achieved by introducing a
regulator function $f_R$ that is $\mathcal{O}(1)$ at low momenta and
$\mathcal{O}(0)$ at high momenta, e.g.
$f_R(p)=\exp\left(-\left(p/\Lambda\right)^n \right)$, where $p$ is the
regulated momentum scale, $\Lambda$ is the cutoff scale that determines
which contributions are discarded, and $n$ is an integer.  Typically,
chiral interactions were regulated nonlocally, but recently, local and
semilocal regularization schemes have been proposed, see e.g.,
References~\cite{Gezerlis:2013ipa,Reinert:2017usi} for more details. 

In this review, we focus on the application of chiral interactions within
QMC methods. Chiral interactions are often constructed in momentum
space and contain various momentum dependencies in terms of the average
incoming and outgoing nucleon momenta $\vb{p}$ and $\vb{p}'$. However,
the GFMC and AFDMC methods are best suited to use local
interactions as input to solve the many-body Schr\"odinger equation,
i.e., the interactions should depend on the momentum transfer
$\vb{q}$. This is true for pion-exchange interactions up to N$^2$LO but not generally true for contact
interactions. However, it is
possible to construct fully local chiral interactions up to N$^2$LO by
choosing a local contact operator basis using Fierz ambiguities and
local regulators. Local chiral interactions have recently been developed
for use in the GFMC and AFDMC methods
within $\Delta$-less~\cite{Gezerlis:2013ipa,Gezerlis:2014zia} and
$\Delta$-full~\cite{Piarulli:2014bda} chiral EFT.

The local regulators for short- and long-range interactions typically are of the form
\begin{align}
\label{eq:regs}
f_{\text{short}}=\alpha \exp\left(-\left(\frac{r}{R_0}\right)^n\right)\,,\quad f_{\text{long}} =\left(1- \exp\left(-\left(\frac{r}{R_0}\right)^{n_1}\right)\right)^{n_2}\,,
\end{align}
where the exponents $n$, ${n_1}$, and ${n_2}$ describe the sharpness of
the regulator and $\alpha$ is a normalization factor.
To cut off the divergences at the origin also
of multi-pion-exchange interactions, the exponents ${n_1}$ and ${n_2}$
need to be chosen sufficiently large.
Note that the minimally nonlocal interactions of
References~\cite{Piarulli:2014bda,Piarulli:2016vel} use
a Woods-Saxon--like functional form for the long-range regulator, which however,
in practice differs only nominally from~\cref{eq:regs}.
More details on the construction
of local chiral interactions can be found in the above-mentioned references.

Nevertheless, local chiral interactions suffer from
regulator artifacts that do not appear for typical nonlocal chiral
interactions. When constructing local contact potentials, Fierz
ambiguities among different contact operators at a given order are
exploited to eliminate nonlocal terms. However, local regulators break
this Fierz rearrangement freedom and short-range regulator artifacts
arise~\cite{Lovato:2011ij,Huth:2017wzw}. It can be shown that the latter are of the
same order as higher-order terms in the chiral expansion and,
therefore, are cured by explicitly including higher-order contact
interactions~\cite{Dyhdalo:2016ygz,Huth:2017wzw}. While this effect
is not very dramatic in the NN sector at, e.g. N$^2$LO, it is quite
sizable in the 3N sector. These
regulator artifacts lead to an ambiguity in the shorter-range $V_D$
and $V_E$ topologies~\cite{Lynn:2015jua}, which increases the theoretical
uncertainty at typical cutoff scales. While local regulators preserve
the analytic structure of the partial-wave amplitude near threshold for
long-range pion-exchange interactions in NN
scattering~\cite{Epelbaum:2014efa}, they also lead to larger regulator
artifacts for the 3N two-pion exchange interaction at typical cutoff
scales; see References~\cite{Tews:2015ufa,Dyhdalo:2016ygz}. Hence, local
chiral interactions together with QMC methods lead to exciting insights,
but one has to carefully consider regulator artifacts that appear and their
influence on the results.

\section{ELECTROWEAK CURRENTS} \label{sec:currents}
The interactions between external electroweak probes -- electrons and neutrinos -- and interacting nuclear systems is described by a set of effective nuclear currents and charge operators. Those associated with neutral-current transitions can be written as~\cite{Shen:2012xz}
\begin{equation}
\label{eq:neutcurr}
J^\mu_{\rm NC}=-2\sin^2\theta_W J^\mu_{\gamma,S} +(1-2\sin^2\theta_W)J^\mu_{\gamma,z} + J^\mu_{5,z}\,,
\end{equation}
where $\theta_W$ is the Weinberg angle ($\sin^2\theta_W = 0.23122$~\cite{Tanabashi:2018oca}), $J^\mu_{\gamma,S}$ and $J^\mu_{\gamma,z}$ are the isoscalar and isovector pieces of the electromagnetic current $J^\mu_\text{EM}=J^\mu_{\gamma,S}+J^\mu_{\gamma,z}$,
and $J^\mu_{5,z}$ denotes the isovector term of the axial current. 

Analogously to the nuclear interaction, electroweak currents can also be expressed as an expansion in many-body operators that act on nucleonic degrees of freedom
\begin{equation}
J^\mu = \sum_i j^\mu(i) + \sum_{i<j} j^\mu(ij)+\cdots
\end{equation}
The one-body charge and current operators have the standard expressions~\cite{Carlson:1997qn} obtained from the nonrelativistic reduction of the covariant single-nucleon current, and include terms proportional up to $1/m^2$, $m$ being the nucleon mass. The transverse ($\perp$) and longitudinal ($\parallel$) components to the momentum transfer $\vb{q}$ of the isoscalar term read
\begin{align}
j^0_{\gamma,S}(i)&= \frac{G_E^S(Q^2)}{2 \sqrt{1+Q^2/(4m^2)}}-\ii\frac{2\, G_M^S(Q^2)-G_E^S(Q^2)}{8m^2}\vb{q}\vdot \left(\vb*{\sigma}_i\cross\vb{p}_i\right)\,, \nonumber\\
\vb{j}^\perp_{\gamma,S}(i)&= \frac{G_E^S(Q^2)}{2m}\vb{p}^\perp_i-\ii\frac{G_M^S(Q^2)}{4m}\vb{q}\cross\vb*{\sigma}_i\,,\nonumber\\
j^\parallel_{\gamma,S}(i)&=\frac{\omega}{q}j^0_{\gamma,S}(i)\,,
\label{eq:jlong}
\end{align}
where $\vb{p}_i$ is the momentum of the $i$th nucleon and current conservation has been used to relate $j^\parallel_{\gamma,S}(i)$ to $j^0_{\gamma,S}(i)$.
The corresponding isovector components of $j^\mu_{\gamma,z}(i)$ are obtained by $G_{E,M}^S(Q^2)\longrightarrow G_{E,M}^V(Q^2)\,\tau_{i,z}$,
with $G_E^{S/V}(Q^2)$ and $G_M^{S/V}(Q^2)$ being the isoscalar/isovector combinations of the proton and neutron electric ($E$) and magnetic ($M$) form factors. 

Omitting for brevity terms proportional to $1/m^2$, the isovector components of the axial weak neutral current $j^{\mu5}_z$ are given by
\begin{equation}
\label{eq:rho5}
j^{0}_{5,z}(i)=-\frac{G_A(Q^2)}{4m}\tau_{i,z}\vb*{\sigma}_i\vdot(\vb{q}+\vb{p}_i)\,, \quad \quad
\vb{j}_{5,z}(i)=-\frac{G_A(Q^2)}{2}\tau_{i,z}\,,
\end{equation}
where $G_A(Q^2)$ is the axial form factor of the nucleon, which is usually parametrized by a dipole
$G_A(Q^2)=g_A/(1+Q^2/M_A^2)$.
The nucleon axial-vector coupling constant is taken to be $g_A = 1.2723$~\cite{Tanabashi:2018oca} and the axial mass $M_A = 1.03$ GeV~\cite{Bernard:2001rs}, as obtained from an analysis of pion electroproduction data~\cite{Amaldi:1979vh} and measurements of the reaction $\nu_\mu+p \to n +\mu$~\cite{Kitagaki:1983px}. Uncertainties in the $Q^2$ dependence of the axial form factor have a significant impact upon neutrino-nucleus cross-section predictions. In particular, the dipole parametrization has been the subject of intense debate: An alternative ``z-expansion'' analyses~\cite{Meyer:2016oeg} has been proposed and dedicated lattice-QCD calculations of $G_A(Q^2)$ have been carried out~\cite{Rajan:2017lxk}.

The charge-changing weak current is written as the sum of polar- and axial-vector components $J^\mu_\text{CC}=J^\mu_{\gamma \pm}+J^\mu_{5,\pm}$,
whose one-body contributions can be obtained from $j^\mu_{\gamma, z}(i)$ and $j^{\mu}_{5,z}(i)$ by replacing  $\tau_{i,z}/2\longrightarrow\tau_{i,\pm} =(\tau_{i,x}\pm\ii\tau_{i,y})/2$.
In addition, one has to retain the induced pseudoscalar contribution~\cite{Marcucci:2001qs,Kaiser:2003dr}
\begin{equation}
j^{\mu}_{5,\text{PS}}(i)=\frac{G_A(Q^2)}{m_\pi^2+Q^2}\tau_{i,\pm}q^\mu\vb*{\sigma}_i\vdot\vb{q}\,.
\end{equation}

The gauge invariance of the theory imposes that the electromagnetic charge and current operators must satisfy the continuity equation $\vb{q}\vdot\vb{j}_\text{EM}=\comm{H}{\rho_\text{EM}}$
where $\rho_\text{EM}\equiv J^0_\text{EM}$, hence providing an explicit
connection between the nuclear interactions and the longitudinal
component of the current operators. For instance, the isospin and
momentum dependence of the NN interactions leads to nonvanishing
commutators with the one-body charge operator and hence to the emergence
of two-body terms
in the current operator. In QMC calculations, both the phenomenological ``Standard Nuclear Physics Approach'' (SNPA) 
and chiral EFT have been exploited to derive many-body current operators.

The SNPA isoscalar and isovector components of the nuclear electromagnetic current $J^\mu_{\gamma,S}$ and $J^\mu_{\gamma,z}$, whose explicit expressions
can be found in Reference~\cite{Shen:2012xz}, lead to a satisfactory description of static properties (charge radii, quadrupole moments, and M1 transition widths),
charge and magnetic form factors of nuclei with $A \leq 12$~\cite{Carlson:1997qn,Marcucci:2005zc,Marcucci:2008mg,Lovato:2013cua}, and electromagnetic 
response functions~\cite{Lovato:2015qka,Lovato:2016gkq}, which will be discussed in~\cref{sec:electroweak}. They consist of ``model-independent'' and ``model-dependent'' terms~\cite{Riska:1989bh}.
The former are obtained from the NN interaction, and by construction satisfy current conservation. The leading operator is the isovector ``$\pi$-like'' current
but important contributions also arise from $\rho$-like terms. The additional two-body currents arising from the momentum-dependence of the NN interaction have 
been numerically proven to be much smaller~\cite{Marcucci:2008mg}.

The transverse components of the two-body currents cannot be directly linked to the nuclear Hamiltonian. In the latest applications of the SNPA formalism~\cite{Shen:2012xz,Lovato:2013cua,Lovato:2015qka,Lovato:2016gkq}, they include the isoscalar $\rho\pi\gamma$ transition and the isovector current associated with the excitation of intermediate $\Delta$-isobar resonances. The $\rho\pi\gamma$  couplings are extracted from the widths of the radiative decay $\rho\to \pi\gamma$~\cite{Berg:1980lwp} and the $Q^2$ dependence of the electromagnetic transition form factor is modeled assuming vector-meson dominance~\cite{Carlson:1991}. Among the model-dependent currents, those associated with the $\Delta$ isobar are the most important ones. 

One of the chief advantages of the chiral EFT formulation is that
electroweak currents are constructed in a consistent fashion with the
nuclear interaction. Since the chiral Lagrangian is
gauge invariant, nuclear electromagnetic currents automatically satisfy
the continuity equation, order by order, with the corresponding chiral
potentials~\cite{Bacca:2014tla}. In particular, an important advantage
of chiral EFT over the SNPA is the explicit connection between the
3N interaction and the two-body axial current. For example, the LEC
$c_D$ entering the 3N potential at N$^2$LO is related to the LEC of the two-body contact
axial current~\cite{Epelbaum:2002vt,Gazit:2008ma}. Similarly to
the EFT interactions discussed in~\cref{sec:interactions}, chiral
EFT currents can also be systematically organized in powers of
$(p/\Lambda_b)^\nu$, where the generic low-momentum scale includes
the momentum transferred by the external electroweak probe. Note
that single-nucleon structure effects have to be accounted for by
introducing appropriate form factors. Because of the shortcomings of 
chiral EFT nucleonic form factors for $Q^2 \gtrsim 0.1$ GeV$^2$~\cite{Walzl:2001vb,Phillips:2003jz}, 
even in chiral EFT formulations, parametrized versions of the latter 
are usually employed.

Over the last decade, extensive work to construct two-body electromagnetic
current operators has been carried out by the JLab-Pisa and
by the Bochum-Bonn groups using standard time-ordered perturbation
theory~\cite{Pastore:2008ui,Pastore:2009is,Pastore:2011ip} and the method
of unitary transformations~\cite{Kolling:2009iq,Kolling:2011mt},
respectively. The LO vector current $\mathbf{j}_{\gamma, S}$ and
$\mathbf{j}_{\gamma, z}$, corresponding to $\nu=-2$, are the same as
those obtained within the SNPA, and are reported in the second and third
lines of~\cref{eq:jlong}.
At N$^2$LO
($\nu=0$) one needs to account for relativistic corrections to the
one-body currents, while at N$^3$LO ($\nu=1$) there are pure two-pion
exchange and short-range one-loop contributions. At this order, additional
``minimal'' and ``nonminimal'' contact diagrams, defined in terms of two
new LECs, need to be accounted for. The former originate from the
contact chiral EFT NN potential at NLO through the minimal substitution
$\vb{p}\to\vb{p}-\ii e\vb{A}$, where $e$ is the electric charge and
$\vb{A}$ is the vector photon field. Consequently, these contributions,
needed for the continuity equation to be satisfied, involve the same
LECs as the contact NN term -- and hence can be determined by fitting
NN scattering data. The nonminimal contributions arise from the field
strengths $F_{\mu\nu}=\partial_\mu A_\nu - \partial_\nu A_\mu $, which
transform covariantly under chiral symmetry~\cite{Pastore:2009is},
and their LECs need to be fixed against electromagnetic observables. It
should be noted that the isoscalar and the isovector contributions of
the minimal terms correspond to the model-dependent $\rho\pi \gamma$
and $\Delta$-excitation transverse currents of the SNPA.

Axial currents were also recently derived within chiral EFT up to
one-loop in References~\cite{Baroni:2015uza,Krebs:2016rqz}, including
pion-pole contributions. The latter are crucial for the current to
be conserved in the chiral limit and are suppressed in low-momentum
transfer processes. In the axial current, OPE contributions enter at
N$^3$LO $(\nu=0)$ and involve the LECs $c_3$, $c_4$, and $c_6$, which
also enter the NN interaction.
The situation is different for the axial charge, as pion-range
contributions already enter at NLO $(\nu=-1)$. One-loop corrections
to the axial current appear at N$^4$LO $(\nu=1)$ and were first
utilized in the calculation of the tritium Gamow-Teller matrix
element~\cite{Baroni:2016xll}. On the other hand, at N$^3$LO $(\nu=1)$
the calculation of the one-loop contributions has been carried out
in Reference~\cite{Baroni:2017gtk}, aimed at studying the inclusive
neutrino scattering off the deuteron at low energies.
Electroweak currents which explicitly include the $\Delta$ excitation,
consistently with the nuclear interactions discussed in~\cref{sec:interactions},
were derived in Reference~\cite{Baroni:2018fdn} and applied to the calculation of the
the Gamow-Teller matrix element contributing to tritium $\beta$ decay.

\section{QUANTUM MONTE CARLO METHODS}\label{sec:QMC}

Quantum Monte Carlo methods provide powerful tools to solve for the ground
state of strongly interacting many-body systems. These
methods have been used for problems in quantum
chemistry and materials with very high accuracy, see e.g.
References~\cite{Hammond:1994,Nightingale:1999,Schmidt:1992,Foulkes:2001}.
Several different QMC implementations exist, both for
bosons and fermions; here we will limit our description to the particular
methods recently used to calculate properties of nuclear systems.

\subsection{Variational Monte Carlo}
The Variational Monte Carlo (VMC) method is used to calculate  observables (e.g. the energy) of a many-body system once a suitable guess for its wave function $\Psi_T$ (the ``trial'' wave function) is provided.\begin{marginnote}[]\entry{VMC}{Variational Monte Carlo}\end{marginnote}
The variational energy $E_V$ of 
an $A$-nucleon system is given by
\begin{equation}
\begin{split}
\label{eq:vmc1}
E_V&=\frac{\ev{H}{\Psi_T}}{\braket{\Psi_T}}
=\frac{\sum_{\sigma \tau}\int\dd\vb{R}
\Psi_T^\ast(\vb{R},\sigma,\tau)
H\Psi_T(\vb{R},\sigma,\tau)}
{\sum_{\sigma\tau}\int\dd\vb{R}
\Psi_T^\ast(\vb{R},\sigma,\tau)
\Psi_T(\vb{R},\sigma,\tau)}\,,
\end{split}
\end{equation}
where $\vb{R}=\{\vb{r}_1\ldots\vb{r}_N\}$, $\sigma=\{\sigma_1\dots\sigma_N\}$, 
and $\tau=\{\tau_1\dots\tau_N\}$ include all particles' positions $\vb{r}_i$, spins $\sigma_i$, and isospins $\tau_i$,
and $H$ is the nuclear Hamiltonian. The energy $E_V$ provides an
upper bound to the ground-state energy $E_0$ and is equal to $E_0$
only if $\Psi_T$ coincides with the true ground-state wave function
of the system, $\ket{\Psi_T}=\ket{\Psi_0}$. The calculation of $E_V$
requires the numerical evaluation of a multidimensional integral, but the
high dimensionality limits standard numerical integration techniques to
very small systems.

Monte Carlo integration is a natural solution to this limitation. \Cref{eq:vmc1} can be rewritten as
\begin{equation}
E_V=\frac{\sum_{\sigma\tau}\int\dd\vb{R}P(\vb{R},\sigma,\tau) 
H\Psi_T(\vb{R},\sigma,\tau)/\Psi_T(\vb{R},\sigma,\tau)}
{\sum_{\sigma\tau}\int\dd\vb{R}P(\vb{R},\sigma,\tau)}\,,
\end{equation}
where the function $P(\vb{R},\sigma,\tau)$ is a probability
distribution, and one natural choice is
$P(\vb{R},\sigma,\tau)=\Psi_T^\dagger(\vb{R},\sigma,\tau)\Psi_T(\vb{R},\sigma,\tau)$. In the VMC method $P$ is used to sample a set of $M$ configurations in $\{\vb{R},\sigma,\tau\}$ space that are used to solve the integral above. A common way to generate such configurations is provided by the Metropolis algorithm, but many others are available. See e.g., Reference~\cite{Foulkes:2001}.

For strongly interacting systems, a common ansatz for variational wave function is
$\ket{\Psi_T}=\hat{F}\ket{\Phi}$.
The correlation operator $\hat{F}$, modeling the short-range correlations 
induced by the Hamiltonian,
can generically be written as
\begin{equation}
\hat{F}=\left(\prod_{i<j} f_c(r_{ij})\right)
\left[\mathcal{S}\prod_{i<j}\left(1+F_{ij}\right)\right]\,,
\end{equation}
where we have omitted three-body correlations just for simplicity. 
In the above equation, $f_c(r)$ is a spin/isospin-independent correlation, and
\begin{equation}
F_{ij}=f_\tau(r_{ij})\tdott{i}{j}+f_\sigma(r_{ij})\sdots{i}{j}
+f_{\sigma\tau}(r_{ij})\sdots{i}{j}\tdott{i}{j}
+f_t(r_{ij})S_{ij}+f_{t\tau}(r_{ij})S_{ij}\tdott{i}{j}\,.
\end{equation}
Evaluating the symmetrization operator $\mathcal{S}$ would require a factorial number of operations. In practice,
the order of pairs for the left and right wave functions is instead
sampled for each configuration. The radial correlations $f_i(r)$ include
variational parameters that are chosen in order to minimize $E_V$.

The long-range antisymmetric part $\ket{\Phi}$ is typically a Slater determinant of single-particle 
orbitals, appropriate for the nuclear system of interest. For homogeneous matter, the orbitals can be 
plane waves or also include pairing correlations~\cite{Gandolfi:2009b}. 
For nuclei, the single-particle orbitals are generally states
written in the $ls$ or $jj$ basis that are properly combined to give
the desired total angular momentum $J$
and isospin $T$ of the nucleus~\cite{Pudliner:1997}. Within the GFMC
method, $\ket{\Phi}$ consists of a set of amplitudes,
each representing a particular spin/isospin configuration of the
many-body state. For example, the spin amplitudes
for 3 neutrons and the amplitudes after a spin/spin operator has acted on the state are represented by:
\begin{align}
\ket{\Phi}=
\left(
\begin{array}{c}
a_{\uparrow\uparrow\uparrow}\\
a_{\uparrow\uparrow\downarrow}\\
a_{\uparrow\downarrow\uparrow}\\
a_{\uparrow\downarrow\downarrow}\\
a_{\downarrow\uparrow\uparrow}\\
a_{\downarrow\uparrow\downarrow}\\
a_{\downarrow\downarrow\uparrow}\\
a_{\downarrow\downarrow\downarrow}
\end{array} \right),
\hspace{6em}\sdots{1}{2}\ket{\Phi}=\left(
\begin{array}{c}
 a_{\uparrow\uparrow\uparrow}\\
 a_{\uparrow\uparrow\downarrow}\\
 2a_{\downarrow\uparrow\uparrow}-a_{\uparrow\downarrow\uparrow}\\
 2a_{\downarrow\uparrow\downarrow}-a_{\uparrow\downarrow\downarrow}\\
 2a_{\uparrow\downarrow\uparrow}-a_{\downarrow\uparrow\uparrow}\\
 2a_{\uparrow\downarrow\downarrow}-a_{\downarrow\uparrow\downarrow}\\
 a_{\downarrow\downarrow\uparrow}\\
 a_{\downarrow\downarrow\downarrow}
\end{array} \right)\,.
\end{align}


The isospin is treated in a similar fashion, except that in this case the number of
elements is smaller due to charge and/or total isospin conservation. For this reason, 
in GFMC calculations the number of many-body spin/isospin states (in the charge basis) 
is equal to $2^A\binom{A}{Z}$. Note that the coefficients associated with a given 
many-body spin/isospin state are such that the wave-function is fully anti-symmetric.


\subsection{Green's Function Monte Carlo}
In the GFMC method, the ground state of the system is obtained with an imaginary-time projection
\begin{equation}
\ket{\Psi_0}\propto\lim_{\tau\to\infty}\exp[-(H-E_0)\tau]\ket{\Psi_T}\,,
\end{equation}
where $\tau$ is the imaginary time, and $E_0$ is a parameter used to control the normalization (that we set to $0$ in the following). The direct computation of the propagator 
$\exp[-H\tau]$ for arbitrary $\tau$ is typically not possible, but for small imaginary times $\delta\tau=\tau/N$ with $N$ large, the calculation is tractable, and the full propagation to large imaginary times $\tau$ can be obtained through the path integral
\begin{equation}
\braket{\vb{R}_N}{\Psi_T}=\int\prod_{i=0}^{N-1}\dd\vb{R}_i\matrixel{\vb{R}_N}{\exp[-H\delta\tau]}{\vb{R}_{N-1}}\cdots\matrixel{\vb{R}_1}{\exp[-H\delta\tau]}{\vb{R}_0}\!\braket{\vb{R}_0}{\Psi_T}\,,
\end{equation}
where Monte Carlo techniques are used to sample the paths ${\vb{R}_i}$. In
practice, a set of configurations, typically called \emph{walkers},
are simultaneously evolved in imaginary time, and then used to calculate
observables once convergence is reached. In the GFMC method, each walker contains
the nucleon positions and a complex amplitude for each spin/isospin state of the
nucleus, implying an unfavorable exponential scaling with the number of nucleons. 

The most common and easiest approximation for the
short-time propagator $G_{\delta\tau}(\vb{R},\vb{R}')\equiv
\matrixel{\vb{R}'}{\exp[-H\delta\tau]}{\vb{R}}$ is obtained by
using the Trotter-Suzuki expansion:
\begin{equation}
\begin{split}
G_{\delta\tau}(\vb{R},\vb{R}')
&=\matrixel{\vb{R}'}{\exp(-V\delta\tau/2)\exp(-T\delta\tau)\exp(-V\delta\tau/2)}{\vb{R}}+\mathcal{O}(\delta\tau^3)\,,
\end{split}
\end{equation}
but more sophisticated and accurate ways to reduce the time-step
error above are available~\cite{Schmidt:1995}.
Here, $T$ is the nonrelativistic kinetic
energy giving rise to the free-particle propagator
$\mel{\vb{R}'}{\exp[-T\delta\tau]}{\vb{R}}\propto\exp[-(\vb{R}-\vb{R}')^2/\lambda^2]$,
with $\lambda^2=4\frac{\hbar^2}{2m}\delta\tau$,
yielding a Gaussian diffusion for the particles.
The matrix $V$ is the spin/isospin-dependent interaction: 
\begin{equation}
\matrixel{\vb{R}}{\exp(-V\delta\tau)}{\vb{R}}
\approx\mathcal{S}\prod_{i<j}\exp[-V_{ij}\delta\tau]\,.
\end{equation}
Each pairwise interaction can be simply evaluated by  exponentiating a
small spin/isospin matrix. This treatment is adequate for static
spin/isospin-dependent NN interactions, as they are diagonal in
coordinate space. In practice one also needs to include
momentum-dependent spin-orbit (LS) NN interactions as well as 3N
interactions. For these and other details, see
Reference~\cite{Carlson:2014vla}.\begin{marginnote}[]\entry{LS}{spin-orbit}\end{marginnote}The
method is exact in the limit of $\delta\tau=0$, and in practice small values of the time step are
used to extrapolate to $\delta\tau=0$.

In addition to ground states, excited states
have been accessed in GFMC calculations. The diffusion
$\lim_{\tau\to\infty}\eu{-H\tau}\ket{\Psi_T}\to\ket{\Psi_0}$ drives
$\ket{\Psi_T}$ to the lowest-energy eigenstate with the same quantum
numbers as $\ket{\Psi_T}$. Thus, to obtain an excited state with distinct quantum numbers from the ground state, one need only construct
a trial wave function with the appropriate quantum numbers. If the excited-state
quantum numbers coincide with the ground state, more care is needed, but results for such states can still be
obtained~\cite{Pieper:2004qw}. 

\subsection{Auxiliary Field Diffusion Monte Carlo}

The basic idea of the AFDMC method~\cite{Schmidt:1999} is to achieve
a better scaling with $A$ than GFMC by sampling the spin/isospin
states rather than explicitly considering all of them. Let us define
the single-nucleon spinor as
\begin{equation}
\label{eq:afdmc_spinor}
\ket{s_i}
=a_i\ket{p\uparrow}+b_i\ket{p\downarrow}+c_i\ket{n\uparrow}+d_i\ket{n\downarrow}\,, \qquad s_i\equiv\{a_i,b_i,c_i,d_i\}
\end{equation}
where $a_i$, $b_i$, $c_i$ and $d_i$ are complex numbers, and $\{\ket{p\uparrow},\ket{p\downarrow},\ket{n\uparrow},
\ket{n\downarrow}\}$ is the proton-up, proton-down, neutron-up and neutron-down basis. The spin/isospin
states employed in AFDMC are products of single-particle states
\begin{equation}
\ket{S} = \ket{s_1} \otimes \dots \otimes \ket{s_A}\, .
\label{eq:sp_states}
\end{equation}
whose dimensionality scales linearly with the number of particles. On the other hand, computing the Slater determinant
of the mean-field part of the wave-function, $\braket{S}{\Phi}$, scales polynomially with the number of nucleons, as it requires 
$A^3$ operations. The main issue associated with the single-particle spin states of~\cref{eq:sp_states} is that they are not closed
with respect to the application of a quadratic spin (or isospin) operators. For instance, it can be easily shown that 
$\sdots{i}{j}\ket{S}\neq\ket{S'}$. For a linear operator instead, one finds
\begin{equation}
\sigma_i^\alpha\ket{S}=\ket{s_1}\otimes \dots \otimes \sigma_i^\alpha \ket{s_i} \otimes \dots \otimes  \ket{s_A}= 
\ket{s_1}\otimes \dots \otimes \ket{s_i^\prime} \otimes \dots \otimes  \ket{s_A}=\ket{S^\prime}
\end{equation}
Realistic nuclear Hamiltonians contain quadratic spin/isospin
operators. Thus, the imaginary-time propagation in AFDMC is carried out using 
the Hubbard-Stratonovich transformation, suitable to linearize
such quadratic spin/isospin dependence:
\begin{equation}
\exp\left(-\frac{1}{2}\lambda\hat{O}^2\right)=\frac{1}{\sqrt{2\pi}}\int \dd x
\exp\left(-\frac{x^2}{2}+\sqrt{-\lambda}x\hat{O}\right)\,,
\end{equation}
where the $x$ are called \emph{auxiliary fields}. For
example, for a spin-dependent interaction,
$v(r_{ij})\sdots{i}{j}\delta\tau$,
it is possible to define new operators such that
$\sigma_i^\alpha\sigma_j^\alpha\to\hat{O}^2$ and
$\lambda\to2v(r_{ij})\delta\tau$~\cite{Carlson:2014vla}.
The strategy of the AFDMC method is to propagate particle positions in the
continuum as commonly done in the GFMC method, but also sample the spin
states of the nucleons in the continuum using the auxiliary fields. For
more details see References~\cite{Sarsa:2003,Carlson:2014vla,Lonardoni:2018prc}.

The AFDMC trial (variational) wave function must be antisymmetric under the exchange of pairs. Omitting again
three-body correlations, it is usually written as 
\begin{equation}
\braket{SR}{\Psi_V} = \bra{SR} 
\left[\prod_{i<j}f_c(r_{ij})\right] \left[1+\sum_{i<j}U_{ij}\right] \ket{\Phi}\,.
\end{equation}
The long range part is given by $\braket{SR}{\Phi} = {\cal A}\{\phi_{\alpha_1}(r_1,s_1)
\dots\phi_{\alpha_N}(r_N,s_N)\}$ with $\phi_{\alpha_i}(\vb{r}_j,s_j)=\braket{\vb{r}_j,s_j}{\phi_{\alpha_i}}=\braket{\vb{r}_j}{f_{n_i}(r)}\braket{s_j}{\xi_i}$
being single-particle orbitals, which are either constructed from a
Wood-Saxon Hamiltonian
or opportunely rescaled from mean-field calculations. While a single
Slater determinant suffices for closed-shell nuclei,
a sum of them is required to described open-shell configurations.
For homogeneous matter, the orbitals can be either plane waves
or pairing correlations.

\section{RESULTS IN LIGHT AND MEDIUM-MASS NUCLEI}
\label{sec:nuclei}

\begin{figure}[t!]
\centerline{
\includegraphics[width=0.5\textwidth]{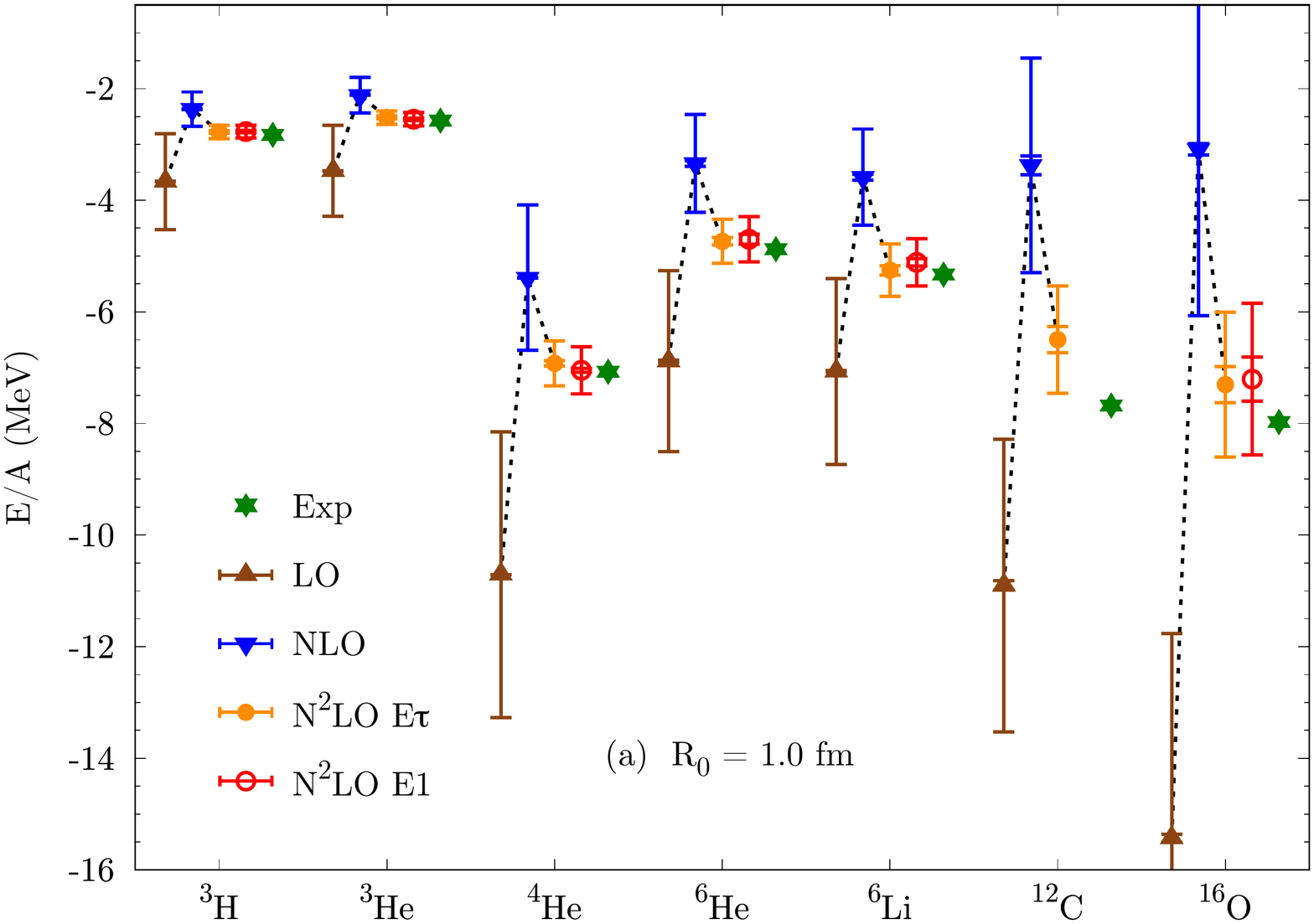}\hspace{-20em}
\includegraphics[width=0.5\textwidth]{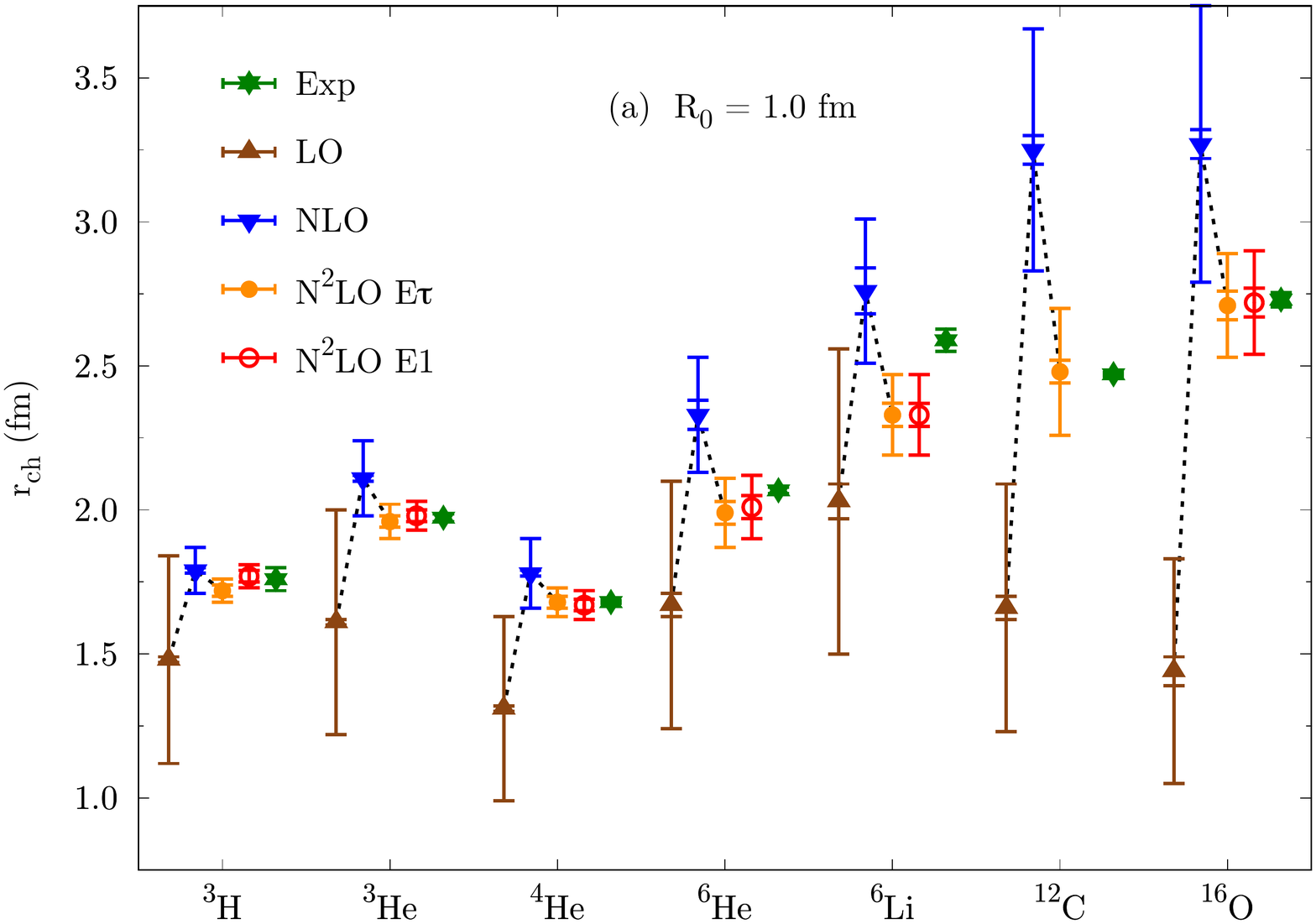}
}
\caption{Ground-state energies (left panel) and charge radii (right panel) for light nuclei with $3\le A\le16$ from AFDMC simulations using NN and 3N chiral EFT interactions at LO (brown upward-facing triangles), NLO (blue downward-facing triangles), and at N$^2$LO with two different parameterizations of the 3N interaction (yellow filled circles and red open circles). The smaller error bars (not always visible) represent the Monte Carlo statistical uncertainties, whereas the larger error bars are  an estimate of the theoretical truncation uncertainty of the chiral expansion. Figure taken from Reference~\cite{Lonardoni:2018prc}.}
\label{fig:afdmceneandrch}
\end{figure}

The combination of \textit{ab initio} QMC methods with interactions
derived from chiral EFT is an exciting development and yields interesting
new insights into nuclear structure. In this section, we present a selection
of recent
results highlighting the rapidly growing reach of this novel combination.

\subsection{Energies, Radii, and Scattering}

One of the most interesting findings that emerge from QMC simulations
with chiral EFT interactions is the ability to simultaneously
describe three different types of nuclear systems: The binding energies and
charge radii of light nuclei, the
LS splitting in the $P$-wave phase shifts of elastic $n$--$\alpha$
scattering, and the equation of state (EOS) of pure neutron matter.\begin{marginnote}[]\entry{EOS}{equation(s) of
state}\end{marginnote}
Historically, these three systems could not be described by a
single combination of phenomenological potentials.
While the combination $\text{AV}18+\text{UIX}$
gives a good description of light nuclei and neutron matter, it
cannot reproduce the LS splitting in the $J^\pi=3/2^-$ and $1/2^-$
partial-wave elastic $n$--$\alpha$ scattering phase shifts. The conclusion
reached was that certain topologies in the 3N interaction (namely
three-pion exchange ring diagrams) were necessary to account for this
LS splitting~\cite{Nollett:2006su}. On the other hand, the combination
$\text{AV}18+\text{IL}7$, which includes such three-pion exchange ring
diagrams, gives an excellent description of light nuclei and the LS splitting
in the $P$ waves of elastic $n$--$\alpha$ scattering but produces a too
soft EOS for neutron matter inconsistent with physical expectations
~\cite{Maris:2013rgq}.
The three-pion-exchange ring diagrams used in the IL7 3N potential appear also in chiral EFT; however, not until N$^3$LO, N$^4$LO, or N$^5$LO (depending on the number of intermediate $\Delta$ states and $\Delta$-full or $\Delta$-less EFT).
Hence, the question emerges how well local chiral N$^2$LO
interactions perform in the three benchmark systems discussed above. We
present results from References~\cite{Lynn:2015jua,Lonardoni:2018prl,Lonardoni:2018prc}
to answer this question.

\begin{figure}[t]
\includegraphics[width=0.7\textwidth,trim=0 1.2cm 0 1.1cm,clip=]{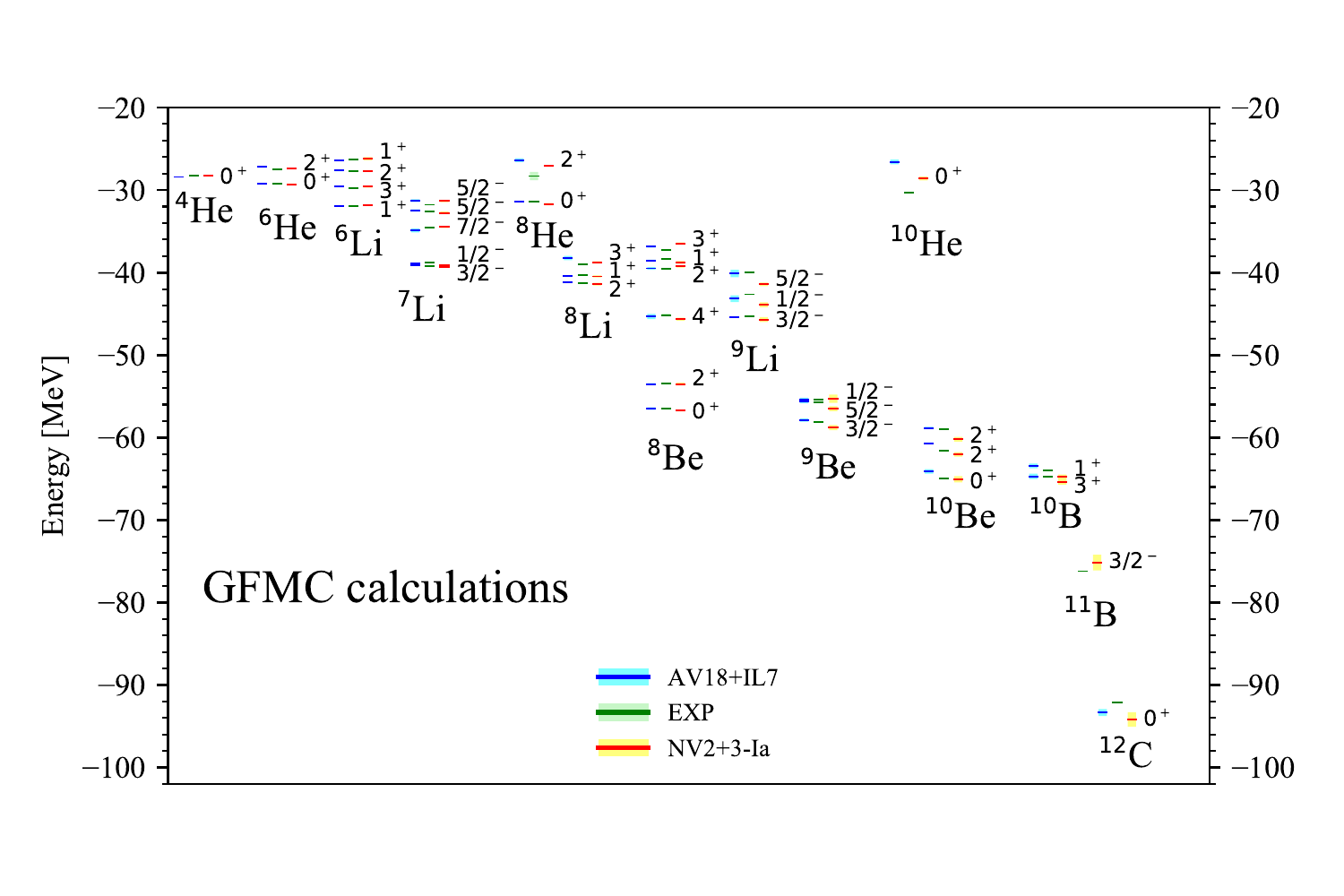}
\caption{Ground- and excited-state energies for light nuclei with $A\le12$ from GFMC calculations using interactions derived from chiral EFT with $\Delta$ degrees of freedom included (red lines) compared with experiment (green lines). Also shown are results using the phenomenological $\text{AV}18+\text{IL}7$ potentials (blue lines). The shaded bands around the lines represent the Monte Carlo statistical uncertainties or the experimental errors. Figure taken from Reference~\cite{Piarulli:2017dwd}.}
\label{fig:gfmc_spectra}
\end{figure}

The binding energies and radii of light nuclei up to \isotope[16]{O}
are presented in~\cref{fig:afdmceneandrch} from AFDMC calculations
reported in Reference~\cite{Lonardoni:2018prl,Lonardoni:2018prc}. The two undetermined LECs
appearing in the N$^2$LO 3N interaction are fit to the \isotope[4]{He}
binding energy and the LS splitting in the $P$-wave phase shifts of
elastic $n$--$\alpha$ scattering. While the agreement for $A=4$ and 5 is by
construction, the very good agreement between
the AFDMC simulations and experiment persists up to \isotope[6]{Li} in
the energies, after which it deteriorates somewhat.
For the radii, the agreement is very good up to \isotope[16]{O}
with an exception for \isotope[6]{Li}, which is also found in simulations
using phenomenological potentials~\cite{Carlson:2014vla}.

The LS splitting between different states is an important feature in light nuclei.  For local chiral EFT interactions at N$^2$LO, in Reference~\cite{Lynn:2015jua} the two 3N LECs were fit to the two $P$-wave phase shifts extracted from an $R$-matrix analysis of the data. While at NLO, the (NN-only) interaction generates too little splitting between the two partial waves (the nonresonant $1/2^-$ partial wave is well reproduced, but not the resonant $3/2^-$ wave), at N$^2$LO with the addition of the 3N interactions, agreement with both partial waves can be reproduced well even for different parameterizations of the 3N interaction.
In Reference~\cite{Piarulli:2017dwd},
ground and some excited states of light nuclei up to \isotope[12]{C}
have been calculated in GFMC using interactions derived from $\Delta$-full
chiral EFT~\cite{Piarulli:2016vel}: See~\cref{fig:gfmc_spectra}. Overall,
the agreement with experiment is very good, with an RMS deviation from
experiment of $<1$~MeV.

\subsection{Distributions and Short-Range Correlations}
\label{subsec:dists}
\begin{figure}
\centerline{
\includegraphics[width=0.5\textwidth]{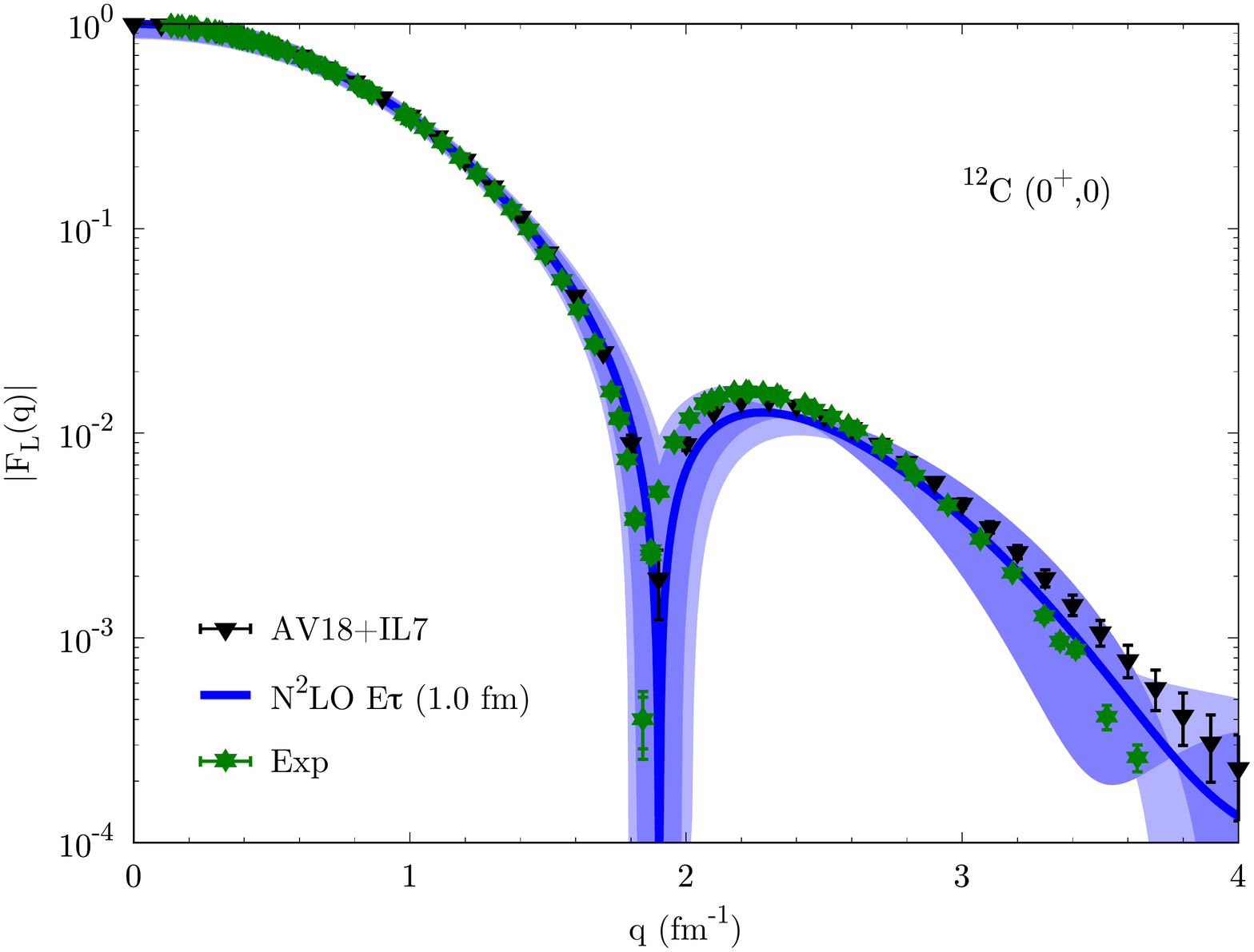}\hspace{-20em}
\includegraphics[width=0.5\textwidth]{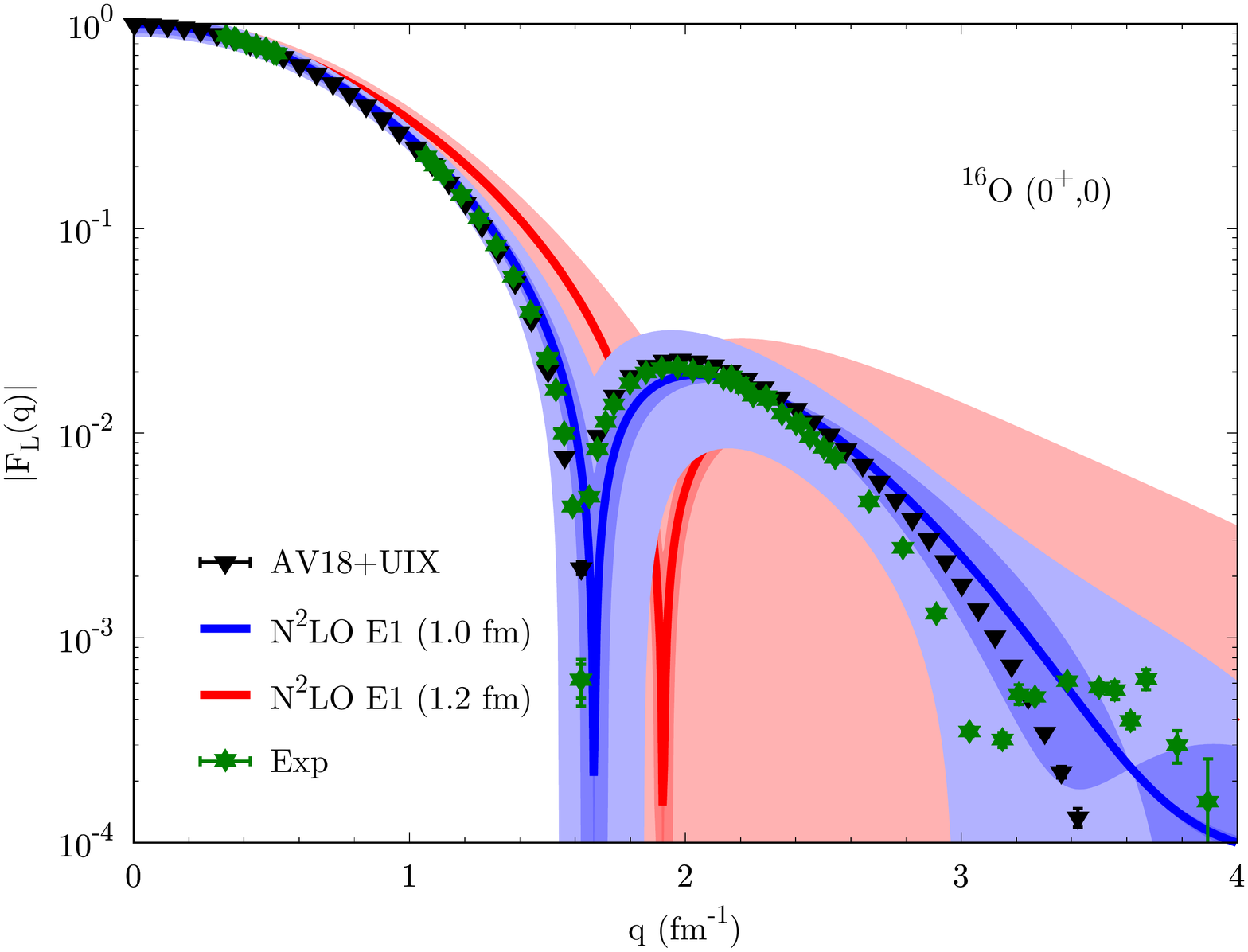}
}
\caption{The longitudinal electric form factors for \isotope[12]{C} (left panel) and \isotope[16]{O} (right panel) from AFDMC calculations using local chiral EFT interactions at N$^2$LO (red and blue bands) compared with experimental data (green circles). For \isotope[12]{C}, only the harder cutoff $R_0=1.0$~fm is shown. Results from GFMC calculations using $\text{AV}18+\text{IL}7$ are shown as the black downward-facing triangles~\cite{Lovato:2013cua}. For \isotope[16]{O}, both cutoffs are shown along with cluster VMC calculations using the $\text{AV}18+\text{UIX}$ potentials (black downward-facing triangles)~\cite{Lonardoni:2017egu}. The bands represent the combined uncertainty coming from the Monte Carlo statistical uncertainties as well as an estimate of the theoretical uncertainty coming from the truncation of the chiral expansion. Figure taken from Reference~\cite{Lonardoni:2018prc}.}
\label{fig:afdmcffs}
\end{figure}

In addition to energies and radii, QMC methods can provide detailed information on the distribution of nucleons in a nucleus in both coordinate and momentum space. These distributions are connected to experimental results in several ways. For example, the one-body point-proton and -neutron densities, defined as
\begin{equation}
\rho_{1,N}(A,r)\equiv\frac{1}{4\pi r^2}\ev{\sum_{i=1}^A\frac{1\pm\tau_{z,i}}{2}\delta(r-|\vb{r}_i-\vb{R}_\text{cm}|)}{\Psi_0},
\end{equation}
with $+$ for the proton ($N=p$) density and $-$  for the neutron ($N=n$) density, are related via Fourier transform to the longitudinal electric form factor $F_L(Q)$; see~\cref{fig:afdmcffs}. Overall, the comparisons of these electric charge form factors with experiment is very good, with the first diffraction minima well reproduced.

\begin{figure}[t]
\includegraphics[width=1.0\textwidth, trim=0 0.1cm 0 0, clip=]{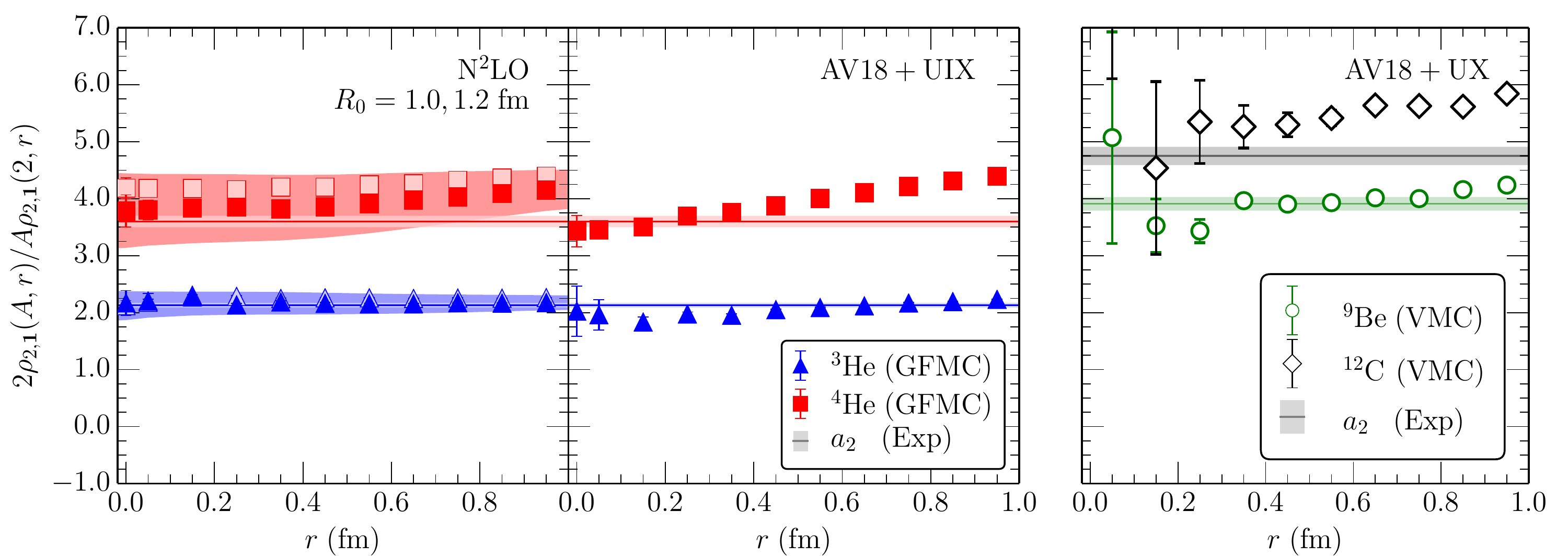}
\caption{Short-range correlation scaling factors obtained from GFMC
(VMC) calculations of light nuclei are shown in the left and middle
panels (right panel) compared with experiment. The left panel shows
results for \isotope[3]{He} and \isotope[4]{He} for local chiral EFT
interactions at N$^2$LO with both a harder ($R_0=1.0$~fm) and a softer
($R_0=1.2$~fm) cutoff. The bands indicate the combined uncertainty
coming from the Monte Carlo statistical errors as well as a theoretical
uncertainty coming from the truncation of the chiral expansion. The
middle panel shows results for the same nuclei with the phenomenological
$\text{AV}18+\text{UIX}$ potentials. The right panel shows results for
\isotope[9]{Be} and \isotope[12]{C} from VMC simulations using the
phenomenological $\text{AV}18+\text{UX}$ potentials. In all panels, the
experimental values with errors are indicated by the horizontal lines
and shaded horizontal regions.} \label{fig:emca2} \end{figure}

Two-body coordinate-space distributions,
\begin{equation}
\rho_{2,\mathcal{O}}(A,r)\equiv\frac{1}{4\pi r^2}\ev{\sum_{i<j}^A\mathcal{O}_{ij}\delta(r-|\vb{r}_{ij}|)}{\Psi_0},
\end{equation}
can also be related to experimentally observable quantities. One of the most interesting results to arise from the novel combination of EFT with QMC methods is the relation of the so-called two-body short-range correlation (SRC) scaling factors $a_2$ to the short-distance behavior of the ratio of the two-body central correlations.\begin{marginnote}[]\entry{SRC}{short-range correlation}\end{marginnote}That is,
\begin{equation}
a_2(A/d)=\lim_{r\to0}\left(2\rho_{2,1}(A,r)/A\rho_{2,1}(2,r)\right),
\end{equation}
with $A$ representing a nucleus with $A$ nucleons and $d$ representing the deuteron ($A=2$); See Reference~\cite{Chen:2016bde} for more details. These SRC scaling factors are extracted from quasielastic scattering from nuclei at intermediate Bjorken $x$ values. In~\cref{fig:emca2} we compare experimental values for $a_2(A/d$) in light nuclei with values extracted from two-body distributions from GFMC calculations using local chiral EFT interactions at N$^2$LO and phenomenological potentials as well as VMC calculations using phenomenological potentials. This novel idea not only sheds new light on two-body SRCs in nuclei, but may also help clarify the nature of the so-far elusive 3N SRCs and on the isospin dependence of the EMC effect through the EMC-SRC linear relationship; see~\cite{Chen:2016bde} for more details.

\begin{figure}[t]
\includegraphics[width=0.58\textwidth, trim=0cm 0.2cm 0 0, clip=]{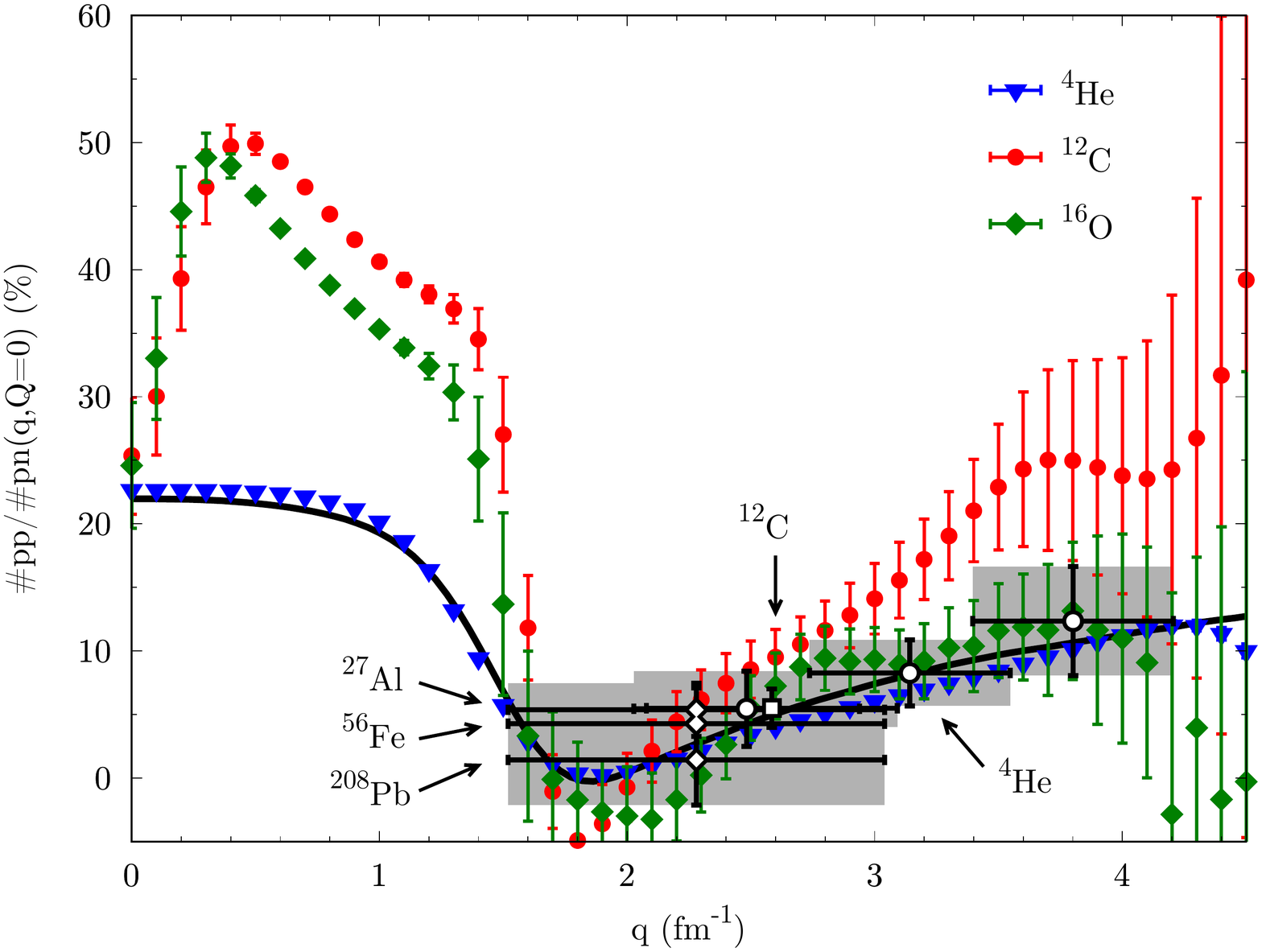}
\caption{Ratio of pp to np pairs in back-to-back kinematics ($\vb{Q}=0$) as a function of the relative momentum $\vb{q}$ between the two nucleons, extracted from two-nucleon momentum distributions using local chiral EFT interactions at N$^2$LO for \isotope[4]{He}, \isotope[12]{C}, and \isotope[16]{O} (blue, red, and green points). The uncertainties shown here are the Monte Carlo statistical uncertainties. Values extracted from experiment are shown as the empty black symbols with gray shaded regions for experimental uncertainties~\cite{Subedi:2008zz,Korover:2014dma,Hen:2014nza}. The solid black line comes from VMC calculations for \isotope[4]{He} using phenomenological potentials ($\text{AV}18+\text{UX}$). Figure taken from Reference~\cite{Lonardoni:2018sqo}.}
\label{fig:pp2nppairs}
\end{figure}

So far, we have discussed distributions calculated in coordinate space (and their Fourier transforms). However, it is also possible to calculate distributions directly in momentum space, such as the two-nucleon momentum distribution (the probability of finding a pair of nucleons in a nucleus with relative momentum $\vb{q}$ and total center-of-mass momentum $\vb{Q}$) 
\begin{equation}
\begin{split}
\rho_\text{NN}(\vb{q},\vb{Q})\equiv\frac{2}{A(A-1)}\sum_{ij}\int\dd\vb{R}\dd\vb{R}'&\Psi^\dagger(\vb{R},\vb{R}')\eu{-\ii\vb{q}\vdot(\vb{r}_{ij}-\vb{r}_{ij}')}\\
&\times\eu{-\ii\vb{Q}\vdot(\vb{R}_{\text{cm},ij}-\vb{R}_{\text{cm},ij}')}\mathcal{P}_\text{NN}(ij)\Psi(\vb{R},\vb{R}')\,,
\end{split}
\end{equation}
where $\mathcal{P}_\text{NN}(ij)=(1/4)(1\pm\tau_{z,i})(1\pm\tau_{z,j})$ is an isospin projector: See Reference~\cite{Lonardoni:2018sqo} for more details.
Such distributions with $\vb{Q}=0$ (so-called ``back-to-back'' pairs)
are valuable to compare with exclusive electron scattering experiments
$^AZ(\text{e},\text{e}'\text{pp})/^AZ(\text{e},\text{e}'\text{np})$,
where the dramatic dominance of np pairs over pp pairs has been
observed~\cite{Subedi:2008zz,Korover:2014dma,Hen:2014nza};
see~\cref{fig:pp2nppairs}. The agreement between the experimentally
extracted ratios and those calculated from two-nucleon momentum
distributions using local chiral EFT interactions at N$^2$LO is very good. In addition, the agreement between the phenomenological results using $\text{AV}18+\text{UX}$ (solid black line in~\cref{fig:pp2nppairs}) is notable.

\section{THE EQUATION OF STATE OF NEUTRON MATTER AND NEUTRON STARS}\label{sec:matter}

The EOS of PNM is closely related to the nuclear symmetry energy that can
be studied at higher densities only in heavy-ion collisions, and to the
EOS of neutron stars (NS), where proton fractions in the core are typically
of the order of only few \%.\begin{marginnote}[]\entry{PNM}{pure neutron
matter}\entry{NS}{neutron
star}\end{marginnote}Therefore, PNM provides a natural bridge between
astrophysical observations of NS and terrestrial nuclear
experiments. Also, the EOS of PNM is an interesting model system to
study nuclear interactions because interactions in PNM are simpler
than in systems containing also protons. For example, only the $T=3/2$
isospin channel contributes to PNM (where $T$ is the total isospin) while
the presence of protons also permits contributions from the $T=1/2$
channel. Nevertheless, the $T=3/2$ isospin channel is only weakly
accessible by studying properties of nuclei.

\begin{figure}[t]
\begin{center}
\includegraphics[width=0.46\textwidth]{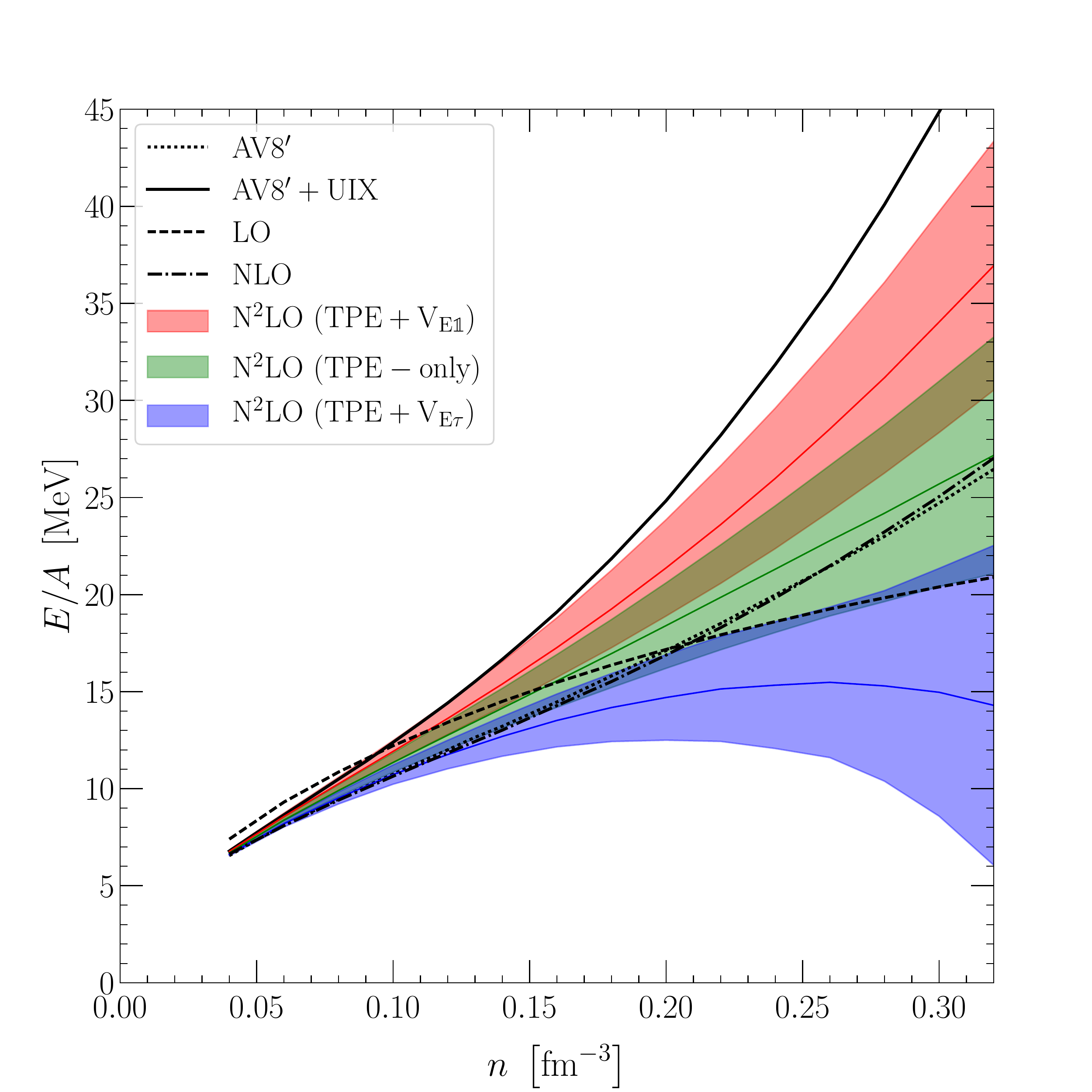}
\end{center}
\caption{The AFDMC EOS of PNM calculated from chiral Hamiltonians at N$^2$LO up to $2n_0$. The different bands correspond to different choices of the 3N short-range operator structure and highlight the impact of regulator artifacts. Each band depicts an uncertainty estimate for the EFT truncation uncertainty. We also show results at LO and NLO as well as results using the phenomenological $\text{AV8}'$ interaction only or also including UIX 3N forces. Figure taken from Reference~\cite{Tews:2018kmu}}
\label{fig:eoschiral}
\end{figure}

While a complete calculation of nuclear matter with arbitrary proton fractions up to $x=0.5$ is still not possible with QMC methods, the AFDMC method has been widely used to calculate the EOS of PNM for many different nuclear interactions in the past years. In practice, in QMC methods the infinite system is simulated by a fixed number of neutrons in a periodic box at a given baryon density. In particular, simulations using 66 neutrons (33 spin up and 33 spin down) give results very close to the thermodynamic limit~\cite{Sarsa:2003zu,Gandolfi:2009fj}. 

In~\cref{fig:eoschiral} we present results for PNM using the AFDMC method
with local chiral interactions up to N$^2$LO. The three different bands
correspond to using different short-range operator structures for the 3N
contact interaction $V_E$ at N$^2$LO as described 
in Reference~\cite{Lynn:2015jua} and discussed in~\cref{sec:interactions}; the differences are due to finite-cutoff
effects and vanish in the limit of large (momentum-space) cutoffs. Each band depicts a
truncation uncertainty estimate based on the order-by-order results
at LO, NLO (both also shown in the figure), and N$^2$LO. The results
are compared to calculations for the phenomenological $\text{AV8}'$ NN
interactions and when additionally including the UIX 3N forces. Note,
that the blue (lower) band produces an EOS that is very soft and leads to
negative pressure at $\approx 1.5 n_0$, which is unphysical. The other
two bands, instead, lead to an EOS that is compatible with calculations
using phenomenological Hamiltonians, but provide uncertainty estimates.

\begin{figure}[t]
\centerline{
\includegraphics[width=0.46\textwidth]{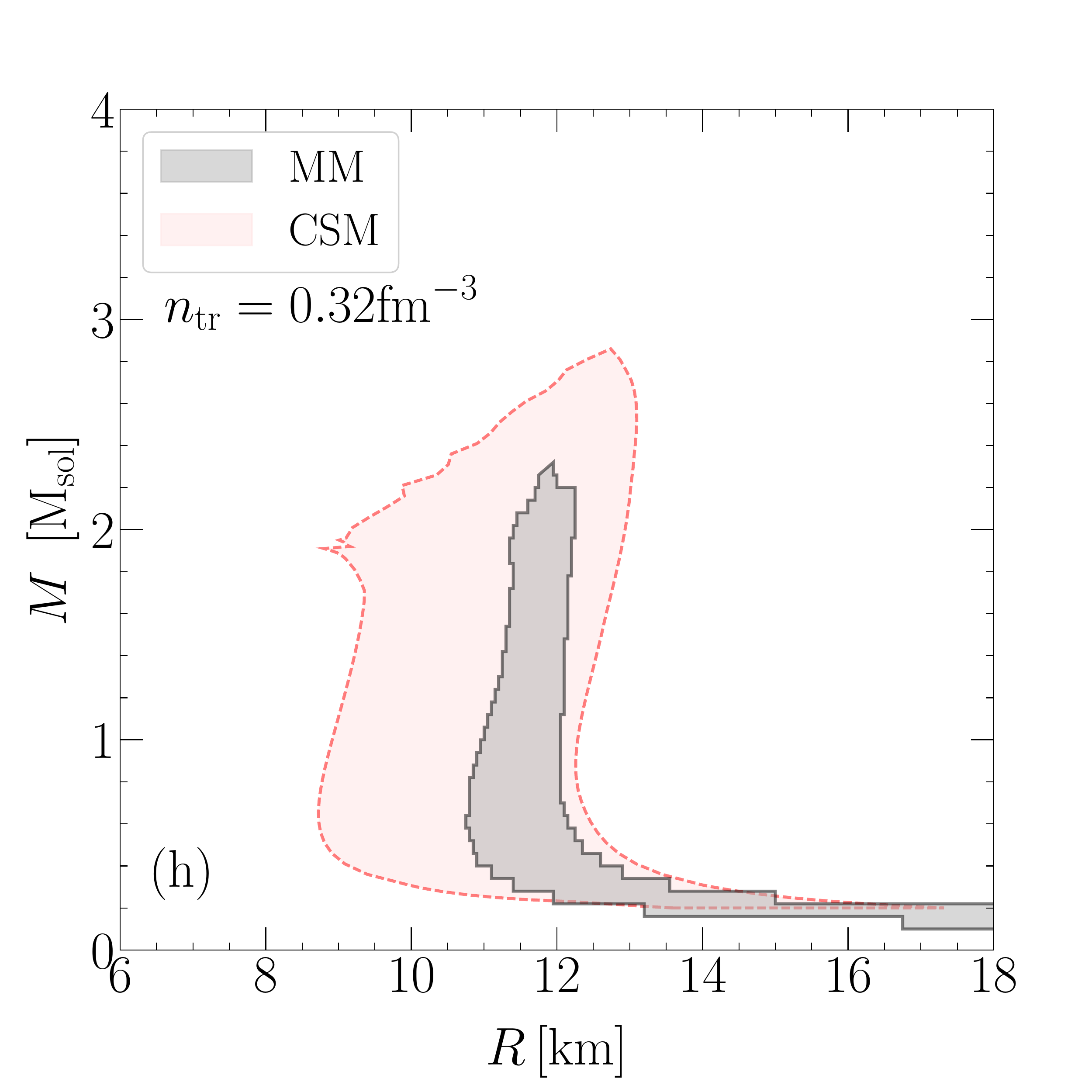}\hspace*{-7cm}
\includegraphics[width=0.45\textwidth]{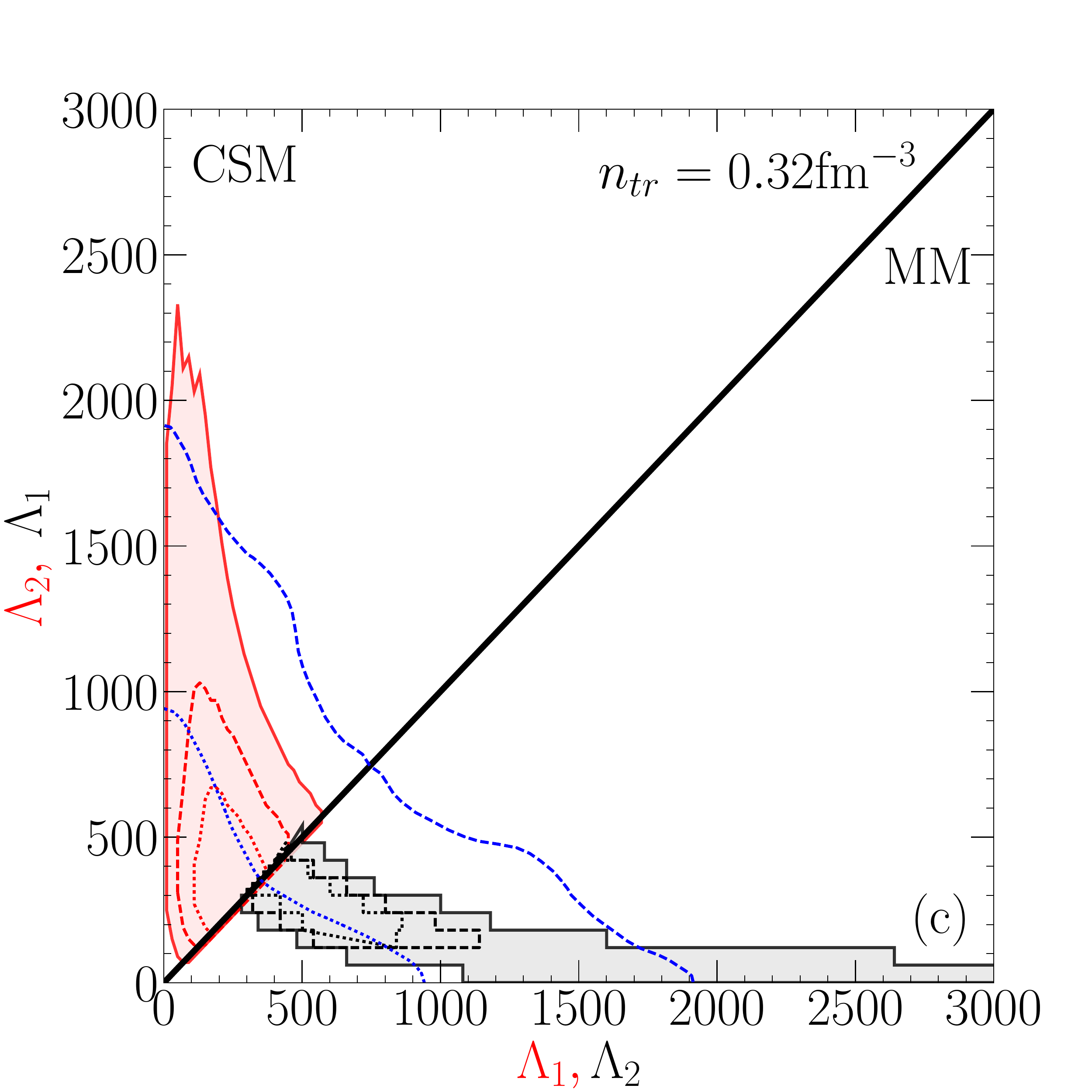}
}
\caption{Left: The mass-radius relation of neutron stars obtained from the QMC EOS up to $2n_0$ and a general extension in the speed of sound for higher densities. The red band (CSM) represents a general extension including also phase transitions while the black band (MM) contains only smooth EOS. 
Right: The corresponding envelopes for the correlation of the tidal polarizabilities $\Lambda_1$ and $\Lambda_2$ of the two neutron stars in the recently observed neutron-star merger GW170817. We also show the 90\% (dashed lines) and 50\% (dotted lines) contours and compare to the corresponding contours from the LIGO-Virgo collaboration (blue). 
The figure is adapted from Reference~\cite{Tews:2018chv}.}
\label{fig:NS_MR_Lam}
\end{figure}

To describe NS, PNM calculations have to be extended to both $\beta$ equilibrium as well as to higher densities. While nuclear Hamiltonians have been used in QMC calculations at all densities encountered in NSs~\cite{Gandolfi:2011xu}, it is not clear if a description in terms of nucleonic degrees of freedom remains valid at high densities. Therefore, a more conservative approach is to use results based on realistic Hamiltonians at small densities, where uncertainties are under control, and extrapolate to higher densities using general extrapolation schemes, e.g. polytropic extensions~\cite{Read:2008iy,Hebeler:2010jx} or speed-of-sound extensions~\cite{Tews:2018kmu,Alford:2015dpa}. With the EOS specified, the structure of an idealized spherically symmetric NS can be calculated by integrating the Tolman-Oppenheimer-Volkoff equations.

Such a general extrapolation has been performed for the PNM EOS from AFDMC calculations in References~\cite{Tews:2018kmu,Tews:2018chv}. In the left panel of~\cref{fig:NS_MR_Lam} we show the resulting mass-radius uncertainty when the QMC input is used up to $2n_0$, for smooth EOS models based on an expansion of nuclear matter around saturation density [minimal model (MM), black band] and for EOS models that are based on a general extension of the speed of sound and also allow phase transitions [speed-of-sound model (CSM), red band]. As one can see, the radius of a typical neutron star has an uncertainty of $\approx 4$~km. While future measurements of neutron star radii from NICER~\cite{NICER} or eXTP~\cite{Watts:2018iom} might strongly constrain the EOS, currently the strongest constraints come from NS mass measurements. Observations in the past eight years~\cite{Demorest:2010,Antoniadis13} have found two NSs with masses near $2M_{\odot}$. These two observations provide some of the strongest constraints on the nature of the EOS above the nuclear saturation density and have been taken into account in~\cref{fig:NS_MR_Lam}.

Finally, we address the recently observed NS merger, GW170817~\cite{TheLIGOScientific:2017qsa,Abbott:2018wiz}. The gravitational-wave signal from this event can be used to constrain the tidal polarizabilities $\Lambda_i$ of the two NSs in the binary system. The tidal polarizability measures to what extent an NS deforms under an external gravitational field, and depends on the compactness of the NS. Using the previously defined EOS, one can compute the tidal polarizabilities of the two NSs in GW170817 and compare the results to the observation by the LIGO-Virgo collaboration~\cite{Tews:2018chv}. We show the results in the right panel of~\cref{fig:NS_MR_Lam} and compare the two extrapolations to the observation (blue). One finds that nuclear physics calculations up to $2n_0$ are more constraining for tidal polarizabilities than the observation of GW170817 with its uncertainty, and that GW170817 does not improve our understanding of the EOS.

\section{RESULTS IN ELECTROWEAK REACTIONS}
\label{sec:electroweak}
The description of neutrino interactions with nuclei 
provides an essential input for current
and planned accelerator-neutrino experiments~\cite{mb_web,
nova_web,t2k_web,dune_web,hk_web}. Since the energy of neutrino beams,
produced as secondary decay products, is not monochromatic, neutrino-oscillation experiments are sensitive to a variety of reaction mechanisms, whose contributions depend on the energy and momentum transfer. The low energy-transfer regime is dominated by coherent scattering, excitations of low-lying nuclear states, and collective modes. At energies of the order of hundreds of MeVs, the leading mechanism is quasielastic scattering, in which the probe interacts primarily with individual bound nucleons. 
Corrections to this leading mechanism arise from processes in which the lepton couples to pairs of interacting nucleons. At higher energies, neutrinos can also excite a struck nucleon to a baryon
resonance state that quickly decays into pions. In this regime, a description solely based on nucleonic degrees of freedom starts to fail, as deep-inelastic scattering (DIS) processes need 
to be accounted for.\begin{marginnote}[]\entry{DIS}{deep inelastic scattering}\end{marginnote}
Achieving a unified description of all these reaction mechanisms is a formidable nuclear theory challenge. Valuable information can be inferred from the analysis of the large wealth of available electron-nucleus scattering data~\cite{Benhar:2006er}. The experiments recently carried out at JLab~\cite{jlab_web} are particularly relevant in this regard, as they probe phenomena occurring at small internuclear distances, such as the reduction in the DIS cross-section ratios for heavier nuclei relative to deuterium (EMC effect), and its connection  to NN SRCs in nuclei~\cite{Chen:2016bde,Hen:2016kwk}; see also~\cref{subsec:dists}.

The interactions of an external electroweak probe with a nucleus are described by the response functions, encoding the strong-interaction dynamics of the nucleons, and their coupling to these external fields. The response functions -- two for the electromagnetic processes, and five for the neutral or charge-changing weak processes -- can be schematically written as
\begin{align}
R_{\alpha\beta}(q,\omega)\!\sim\! &\sum_f \mel{f}{J_\alpha(\omega,\mathbf{q})}{0}^*\!\mel{f}{J_\beta(\omega,\mathbf{q})}{0}\delta(\omega+E_0-E_f)
\label{eq:respon_def}
\end{align}
where ${\bf q}$ and $\omega$ are the momentum and energy transfers
injected by the external field into the nucleus, $\vert 0\rangle$ and $\vert f\rangle$ represent respectively its initial ground state of energy $E_0$ and final states of energy $E_f$ (possibly in the continuum), and $J_{\alpha}$ denotes the appropriate components of the nuclear electroweak current operator~\cite{Shen:2012xz,Rocco:2018mwt}.

Even at intermediate values of the momentum transfer ($q \lesssim 0.5$ GeV) and for energy-transfer corresponding to the quasielastic region, the calculation of the response functions involves severe difficulties, as it requires summation over the entire excitation spectrum of the nucleus and the inclusion of one- and many-body terms in the electroweak currents. 

Integral properties of the responses can be studied by means of their sum rules 
\begin{equation}
S_{\alpha\beta}(q)=C_{\alpha\beta}(q) \int \dd\omega\, R_{\alpha\beta}(q,\omega)\,,
\label{eq:sr_def}
\end{equation}
where  $C_{\alpha\beta}$ are $q$-dependent normalization factors. 
Fixing the $\omega$-dependence of the current operators at the quasielastic peak, $\omega_\text{qe}=\sqrt{q^2+m^2}-m$,
the sum rules can be expressed as ground-state expectation values $S_{\alpha\beta}(q)=\ev{J_\alpha^\dagger(\omega_\text{qe},\vb{q})J_\beta(\omega_\text{qe},\vb{q})}{0}$.

GFMC calculations of the sum rules of the electromagnetic and neutral-current response functions of \isotope[12]{C} have been reported in 
References~\cite{Lovato:2013cua,Lovato:2014eva}. Processes involving two-body currents substantially increase ($\simeq 30$\%) the one-body sum rules 
even down to small momentum transfers. At low momentum transfers terms involving two-body currents only dominate, primarily with the same pair contributing in both
$J_\alpha^\dagger$ and $J_\beta$. 
At higher momentum transfers the interference between one- and two-nucleon currents plays a more important role, as noted
in Reference~\cite{Benhar:2015ula}. Consistently with the analysis carried out by experiments at Brookhaven National Laboratory~\cite{Piasetzky:2006ai} and JLab~\cite{Subedi:2008zz,Hen:2014nza} on exclusive 
measurements of back-to-back pairs in \isotope[12]{C}, the contribution of np pairs is the most important one. This has to be ascribed to the 
tensor component of the nuclear interaction, which plays a larger role in np pairs where it can act in relative $S$-waves, while it acts 
only in relative $P$-waves  (and higher partial waves) in nn and pp pairs~\cite{Schiavilla:2006xx,Alvioli:2007zz,Wiringa:2008dn,Weiss:2018tbu}. 

The sum rules calculations are not capable of 
identifying the energy-transfer dependence of the calculated
excess strength induced by two-body currents. A direct GFMC calculation
of $R_{\alpha\beta}(q,\omega)$ is impractical, because it would require
evaluating each individual transition amplitude $\ket{0}\longrightarrow
\ket{f}$ induced by the current operators.  To circumvent this
difficulty, the use of integral transform techniques has proved to be
quite helpful~\cite{Orlandini:2016hsk}. Such an approach is based on
the Laplace transform of
the response functions, i.e. the Euclidean
response~\cite{Carlson:1992ga,Carlson:2001mp}, defined as

\begin{equation}
E_{\alpha\beta}(q,\tau) = C_{\alpha\beta}(q)\int_{\omega_{\rm el}}^\infty\dd\omega\,
 \eu{-\tau \omega} R_{\alpha\beta}(q,\omega) \ .
\label{eq:laplace_def}
\end{equation}
The lower integration limit $\omega_{\rm el} = q^2/2M_A$, $M_A$ being the mass of the target nucleus, is the elastic scattering
threshold -- corresponding to the $\ket{f}=\ket{0}$ term in the sum of~\cref{eq:respon_def} -- whose contribution is excluded.
Using the same procedure as in the sum-rule calculations to fix the $\omega$ dependence of the current operators, the Euclidean responses
can be expressed as ground-state expectation values, 
\begin{equation}
\frac{E_{\alpha\beta}(q,\tau)}{C_{\alpha\beta}(q)}= \frac{\ev{J^\dagger_\alpha(\omega_\text{qe},\vb{q})\eu{-(H-E_0)\tau}J_\beta(\omega_\text{qe},\vb{q})}{0}}{\ev{\eu{-(H-E_0)\tau}}{0}}\,.
\label{eq:euc_me}
\end{equation}
The Euclidean response functions reported in References~\cite{Lovato:2015qka,Lovato:2016gkq,Lovato:2017cux} are computed with the variational wave function, $\ket{0}=\ket{\Psi_V}$. This is justified by the fact that the sum rules computed with $\ket{\Psi_V}$ for \isotope[12]{C} are very close (within less than $5\%$) to those computed with the GFMC wave function~\cite{Lovato:2013cua,Lovato:2014eva}. The calculation of the matrix element above proceeds in two steps~\cite{Carlson:1987}. First, an unconstrained imaginary-time propagation of the VMC state $\ket{\Psi_V}$ is performed and saved. Next, the states $J_\beta(\omega_\text{qe},\vb{q}) \ket{\Psi_V}$ are evolved in imaginary time following the path previously saved. During this latter imaginary-time evolution, scalar products of $\exp\left[-\left(H-E_0\right)\tau_i\right]J_{\beta}(\omega_\text{qe},\vb{q}) \ket{\Psi_V}$ with $J_\alpha(\omega_\text{qe},\vb{q})\ket{\Psi_V}$ are evaluated on a grid of $\tau_i$ values, and from these scalar products estimates for $E_{\alpha\beta}(q,\tau_i)$ and for the associated statistical error are obtained~\cite{Carlson:1992ga,Carlson:2001mp}.

\begin{figure}
\centerline{
\includegraphics[width=0.48\textwidth]{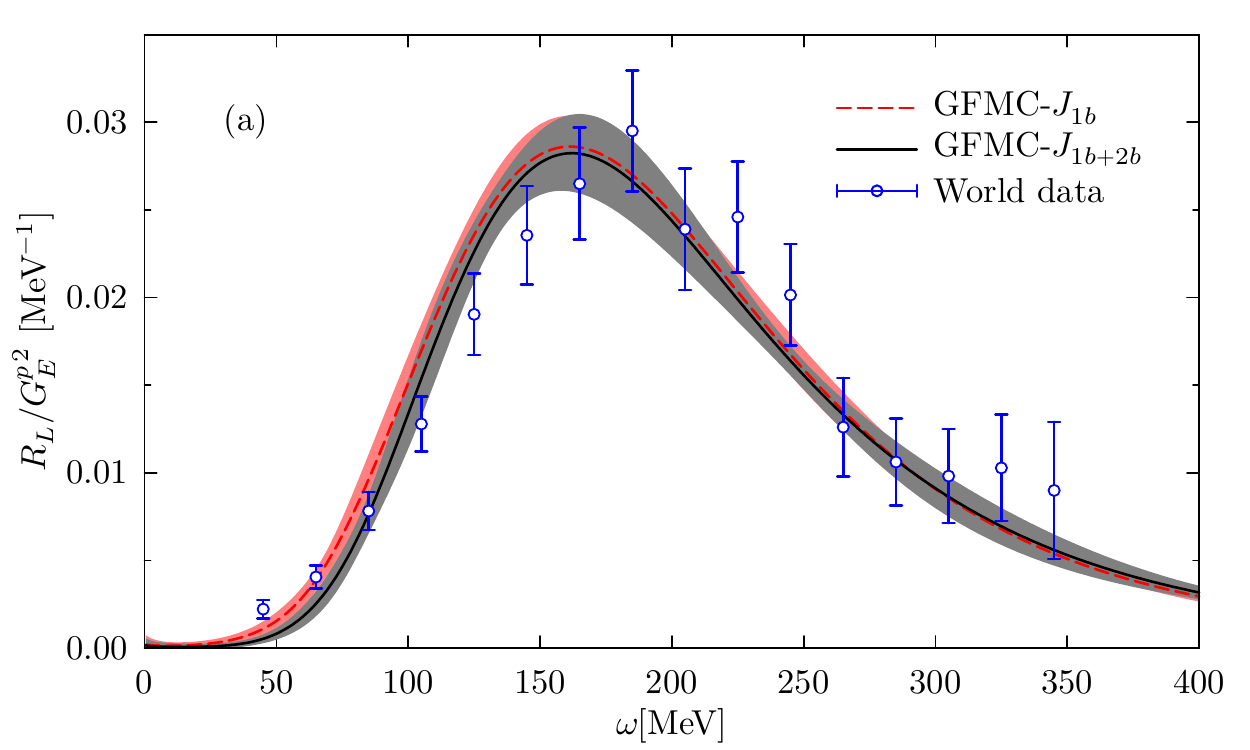}\hspace{-6.8cm}
\includegraphics[width=0.48\textwidth]{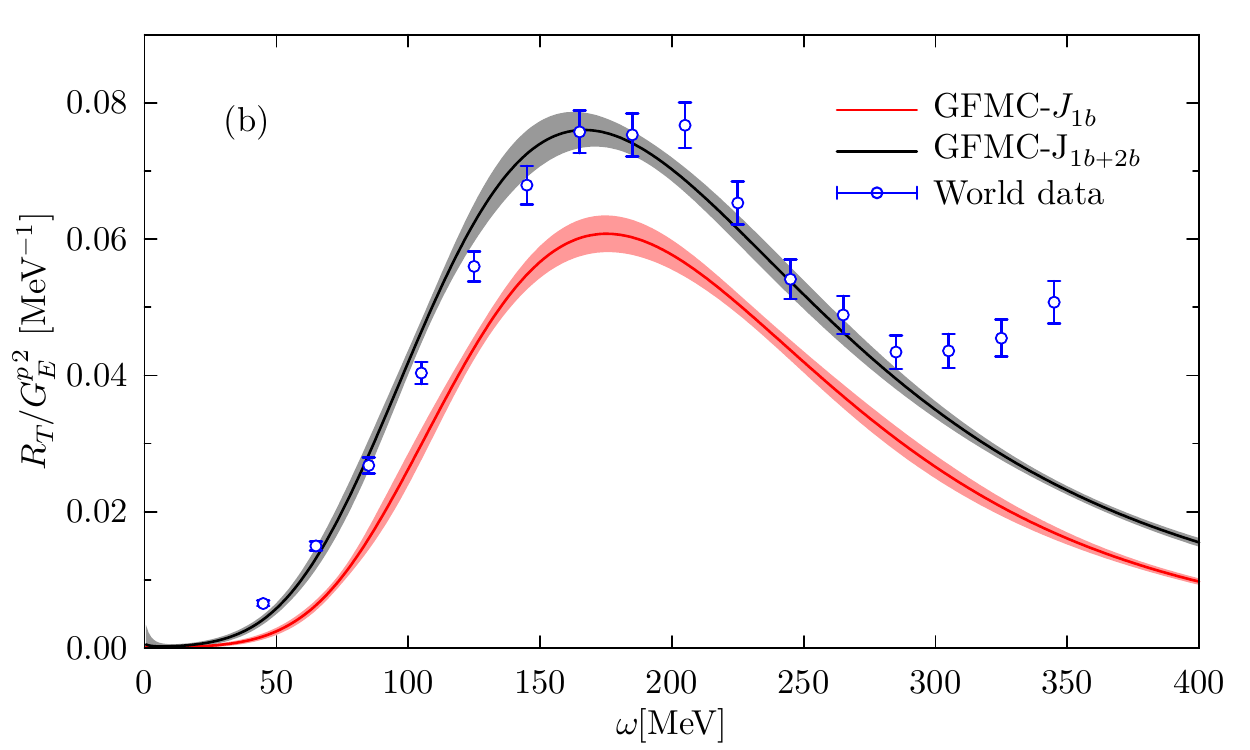}
}
\centerline{
\includegraphics[width=0.48\textwidth]{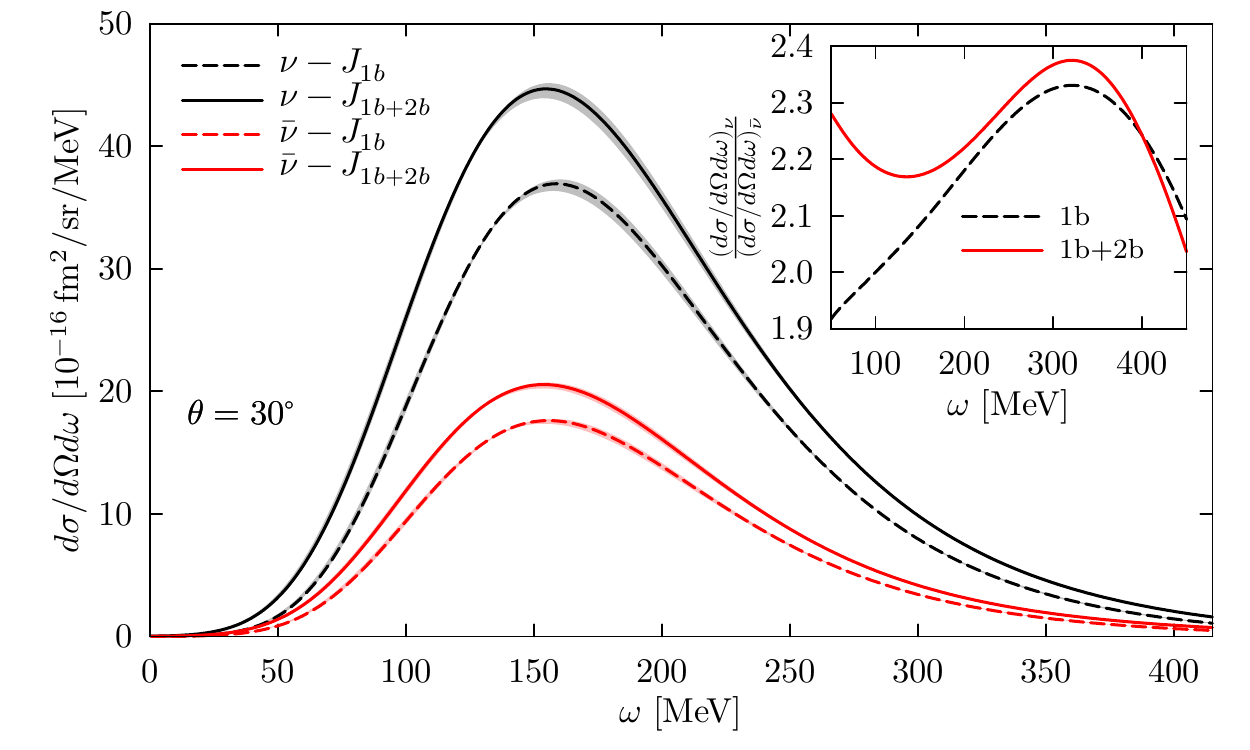}\hspace{-6.8cm}
\includegraphics[width=0.48\textwidth]{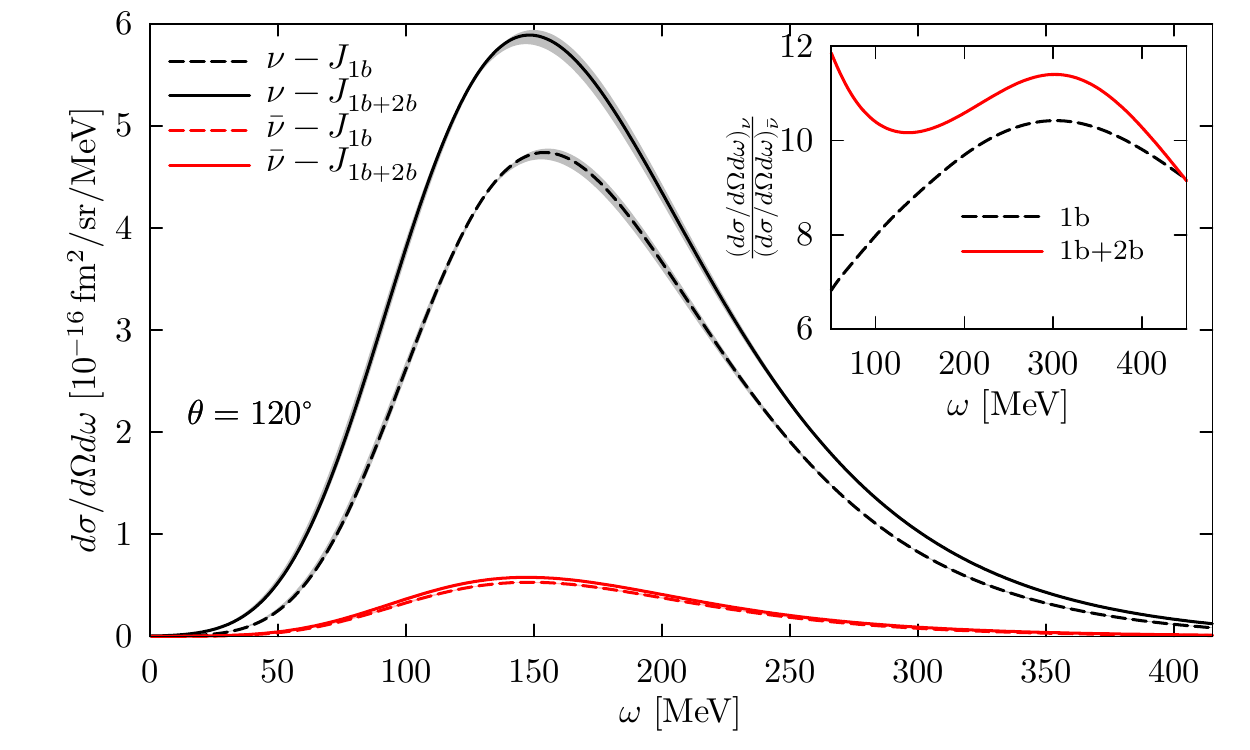}
}
\caption{(Top two panels) Electromagnetic longitudinal (left panel) and transverse (right panel) response functions of \isotope[12]{C} for $q=570$ MeV obtained with one-body only (red dashed line) and one- and two-body (black solid line) terms in the electromagnetic current. Experimental data are from References~\cite{Jourdan:1996np}. (Bottom two panels) Weak neutral $\nu$ (black curves) and $\bar{\nu}$ (red curves) differential cross sections of $^{12}$C at $q = 570$ MeV/c, obtained with one-body only and one- and two-body terms in the neutral current operator, for final neutrino angle $\theta=30^{\circ}$ (left panel) and $\theta=120^{\circ}$ (right panel). The insets show ratios of the $\nu$ to $\bar{\nu}$ cross sections. The figure is adapted from Reference~\cite{Lovato:2017cux} (top two panels) and Reference~\cite{Lovato:2016gkq} (bottom two panels).}
\label{fig:em_response_nc_xsec}
\end{figure}

Retrieving the energy dependence of the response functions requires a numerical inversion of the Laplace transform of~\cref{eq:laplace_def}, a notoriously ill-posed problem. The GFMC calculations of the electroweak response functions carried out over the last few years exploit maximum entropy techniques~\cite{Bryan1990,Jarrell:1996rrw} to perform the analytic continuation of the Euclidean response function. More specifically, the so called ``historic maximum entropy'' technique, employed in Reference~\cite{Lovato:2015qka}, has been augmented to better propagate the statistical errors associated with $E_{\alpha\beta}(q,\tau)$. 
By exploiting GFMC and maximum entropy techniques, the authors of Reference~\cite{Lovato:2016gkq} have demonstrated that accurate calculations of the response, based on a realistic correlated nuclear wave function and containing one- and two-body currents, can reproduce the \isotope[12]{C} electromagnetic response functions in the quasielastic region. In the top two panels of~\cref{fig:em_response_nc_xsec}, the GFMC response functions of \isotope[12]{C} at $q=570$ MeV in which only one-body or both one- and two-body terms are included in the electromagnetic current operators -- denoted by (red) dashed and (black) solid lines and  labeled GFMC-J$_{1b}$ and GFMC-J$_{1b+2b}$, respectively -- are compared to the experimental world data analysis of Reference~\cite{Jourdan:1996np}. The red and gray shaded areas show the uncertainty of the inversion procedure, ultimately associated with the statistical error of the corresponding Euclidean responses. While the contributions from two-body charge operators tend to slightly reduce the longitudinal response in the threshold region, those from two-body currents generate a large excess of strength in the transverse channel, significantly improving the agreement with experimental data. The absence of explicit pion production mechanisms restricts the applicability of the GFMC method to the quasielastic region of the transverse response. Within this picture, the so-called quenching of the longitudinal response near the quasielastic peak emerges as a result of initial- state correlations and final-state interactions, as opposed to the in-medium modification of the nucleon form factors advocated in Reference~\cite{Cloet:2015tha}.

The $\nu$ and $\overline\nu$ differential cross sections and the $\nu / \overline\nu$ ratios for a fixed value of the three-momentum transfer ($q\,$=$\,$570 MeV/c) as function of the energy transfer for a number of scattering angles are displayed in the bottom two panels of~\cref{fig:em_response_nc_xsec}. In terms of the response functions, they are given by
\begin{align}
\frac{\dd\sigma}{\dd\omega \, \dd\Omega} =& \frac{G_F^2}{2\pi^2} k^\prime E^\prime \cos^2\frac{\theta}{2}
 \Bigg[ R_{00}(q,\omega)+
\frac{\omega^2}{q^2}R_{zz}(q,\omega)
-\frac{\omega}{q}R_{0z}(q,\omega)\nonumber\\
&+ \Big(\tan^2\frac{\theta}{2} + \frac{Q^2}{2\,q^2}\Big)R_{xx}(q,\omega)
\mp \tan\frac{\theta}{2} \sqrt{\tan^2\frac{\theta}{2}+
\frac{Q^2}{q^2}}R_{xy}(q,\omega)\Bigg]\ ,
\label{eq:nc_xs}
\end{align}
where $-$ ($+$) refers to $\nu$ ($\overline\nu$), $k^\prime$ and $E^\prime$ are the momentum and energy of the outgoing neutrino, $q$ and $\omega$ 
are the momentum and energy transfers with $Q^2=q^2 -\omega^2$ being the four-momentum transfer, $\theta$ is the outgoing neutrino scattering angle relative
to the direction of the incident neutrino beam, and $G_F=1.1803\times 10^{-5}$ GeV$^{-2}$~\cite{Towner:1999}.
GFMC results for the response functions and cross sections relevant to neutrino scattering off $^{12}$C induced by neutral-current transitions are reported in Reference~\cite{Lovato:2017cux}. These calculations are based on the same dynamical model employed in the electromagnetic case: the nucleons interact with each other via AV18 and IL7 nuclear potentials and with electroweak fields via the phenomenological currents presented is~\cref{sec:currents}. 
Because of the cancellation between the $R_{xx}$ and $R_{xy}$ response functions in~\cref{eq:nc_xs}, the $\overline{\nu}$ cross section decreases rapidly relative to the $\nu$ one as the scattering angle changes
from the forward to the backward hemisphere.  For analogous reasons, two-body currents, which mostly impact the $R_{xx}$ and $R_{xy}$ responses, are larger for the $\nu$ than for the $\overline{\nu}$ cross section, becoming almost negligible in the latter case for backward-angle cross section. Their contributions significantly increase the magnitude of the cross sections over the entire quasielastic region, and particularly the ratio of neutrino to anti-neutrino cross sections. The analysis of the five response functions entering~\cref{eq:nc_xs} reveals that this enhancement is mostly due to constructive interference between the one- and two-body current matrix elements, and is consistent with that expected on the basis of sum-rule analyses discussed earlier. It has to be noted that, at variance to the electromagnetic case, two-body terms in the weak neutral charge also produce excess strength in $R_{00}$ and $R_{0z}$ beyond the quasielastic peak.

A major limitation of the GFMC response functions comes from
the nonrelativistic nature of the calculation. In
Reference~\cite{Rocco:2018tes} the applicability of GFMC has been extended 
in the
quasielastic region to intermediate momentum transfers by performing the
calculations in a reference frame that minimizes the momenta of the struck
nucleon. Additional relativistic effects in the kinematics are accounted
for employing the two-fragment model, which relies on the assumption that
the quasielastic reaction is dominated by the break-up of the nucleus
into a knocked-out nucleon and a remaining $(A-1)$
system. This assumption enables one to connect, in a relativistically
correct way, the energy transfer to the excitation energy of the nucleus,
entering the energy-conserving delta function of the scattering process.
It has to be noted that the two-fragment model has been adopted only
for determining the kinematic input of a calculation where the full
nuclear dynamics of the system is taken into account. This is achieved by
interpolating the GFMC response function at energy and momentum transfer
that fulfill a relativistic energy-conserving delta function. Despite the
two-fragment model does not contain tunable parameters, it 
improves the agreement between experimental data and GFMC calculations for
the inclusive electron-$^4$He cross sections, especially for relatively
large values of the incoming lepton energy -- see \textbf{Figure 7}
of Reference~\cite{Rocco:2018tes}.

\section{SUMMARY AND OUTLOOK}

In this review we have presented recent advances in QMC methods for
nuclear physics. Most of these advances have been made possible both by
new developments of many-body methods themselves and by the
implementation of systematic interactions from chiral EFT. We have presented 
results showing that this fruitful combination can accurately describe energies of
ground and excited states, radii, momentum distributions of nuclei up to
\isotope[16]{O}, and $n$-$\alpha$ scattering. At the same time, QMC
methods with chiral interactions give a reliable description of neutron
matter for astrophysical applications, such as neutron stars and
neutron-star mergers. We have also reviewed exciting results on
electroweak reactions, which are important for the understanding of
interactions between neutrinos and matter. 

Looking forward to the future, there is still interesting and important
work to be done.  A remaining milestone is the combination of accurate
QMC methods with consistent chiral EFT interactions and electroweak
currents and the study of heavier systems with this combined approach.
To achieve these goals, both QMC methods and local chiral EFT
interactions need to be improved.

For the methods, one question that must be addressed is how to build a
wave function for AFDMC calculations of nuclei, which is sophisticated
enough to capture important correlations for larger nuclear systems such
as \isotope[40]{Ca}, while still maintaining favorable scaling in $A$.
In addition, the implementation of different boundary conditions might
allow one to access the thermodynamic limit in nuclear matter for
smaller particle numbers, which might permit systematic computations of
asymmetric matter.  
On the interaction side, extending minimally nonlocal chiral
interactions to higher orders in the EFT power counting, including
consistent many-body forces, will be an important step forward. For
example, complete chiral interactions at N$^3$LO will help to reduce the
systematic uncertainties and allow to determine how well the chiral
expansion is converging in the local chiral EFT approach. Within this
context, it would be desirable to make a systematic comparison, order by
order, of the $\Delta$-less and $\Delta$-full local chiral interactions to better understand the effect of $\Delta$ degrees of
freedom on the order-by-order convergence.  The effect of
regulator artifacts and a wider range of cutoffs also needs to be studied in
few- and many-body systems. Finally, explicitly including pion fields in
QMC methods~\cite{Madeira:2018ykd} might offer new insights on the
chiral expansion.


\section*{DISCLOSURE STATEMENT}
The authors are not aware of any affiliations, memberships, funding, or financial holdings that might be perceived as affecting the objectivity of this review. 

\section*{ACKNOWLEDGMENTS}
We wish to give our special thanks to the late Steven C. Pieper, who
contributed so much to the progress of nuclear QMC methods. We also thank J. Carlson, D. Lonardoni, R. Schiavilla, A. Schwenk, and
R.B. Wiringa.
The work of J.E.L. was supported by the BMBF under contract No
05P15RDFN1. 
The work of I.T. and S.G. was supported by the U.S. DOE under contract
DE-AC52-06NA25396, and by the LANL LDRD program. S.G. was also supported
by the NUCLEI SciDAC program and by
the DOE Early Career Research Program.
The work of A.L. was supported by the U.S. DOE under contract
DE-AC02-06CH11357.
This research used resources provided by the Los Alamos National
Laboratory Institutional Computing Program, which is supported by the
U.S. Department of Energy National Nuclear Security Administration under
Contract No. 89233218CNA000001.
Computational resources have been provided by the National
Energy Research Scientific Computing Center (NERSC), which is supported
by the U.S. Department of Energy, Office of Science, under Contract
No. DE-AC02-05CH11231. Calculations for this research were also conducted
on the Lichtenberg high performance computer of the TU Darmstadt.

%

\bibliographystyle{ar-style5.bst}
\bibliography{annrevqmc}

\begin{thebibliography}{143}
\expandafter\ifx\csname natexlab\endcsname\relax\def\natexlab#1{#1}\fi

\bibitem{Simonis:2017dny}
Simonis J, et~al.
\newblock \textit{Phys. Rev. C} 96:014303 (2017)

\bibitem{Hergert:2015awm}
Hergert H, et~al.
\newblock \textit{Phys. Rept.} 621:165 (2016)

\bibitem{Leistenschneider:2017mrr}
Leistenschneider E, et~al.
\newblock \textit{Phys. Rev. Lett.} 120:062503 (2018)

\bibitem{Hagen:2016uwj}
Hagen G, Jansen GR, Papenbrock T.
\newblock \textit{Phys. Rev. Lett.} 117:172501 (2016)

\bibitem{Morris:2017vxi}
Morris TD, et~al.
\newblock \textit{Phys. Rev. Lett.} 120:152503 (2018)

\bibitem{Pastore:2017uwc}
Pastore S, et~al.
\newblock \textit{Phys. Rev. C} 97:022501 (2018)

\bibitem{Gandolfi:2012}
{Gandolfi} S, {Carlson} J, {Reddy} S.
\newblock \textit{Phys. Rev. C} 85:032801 (2012)

\bibitem{Tews:2018kmu}
Tews I, Carlson J, Gandolfi S, Reddy S.
\newblock \textit{Astrophys. J.} 860:149 (2018)

\bibitem{Bedaque:2002mn}
Bedaque PF, van Kolck U.
\newblock \textit{Ann. Rev. Nucl. Part. Sci.} 52:339 (2002)

\bibitem{Epelbaum:2008ga}
Epelbaum E, Hammer HW, Mei{\ss}ner UG.
\newblock \textit{Rev. Mod. Phys.} 81:1773 (2009)

\bibitem{Machleidt:2011zz}
Machleidt R, Entem DR.
\newblock \textit{Phys. Rep.} 503:1 (2011)

\bibitem{Carlson:2014vla}
Carlson J, et~al.
\newblock \textit{Rev. Mod. Phys.} 87:1067 (2015)

\bibitem{Epelbaum:2011md}
Epelbaum E, Krebs H, Lee D, Meissner UG.
\newblock \textit{Phys. Rev. Lett.} 106:192501 (2011)

\bibitem{Savage:2011xk}
Savage MJ.
\newblock \textit{Prog. Part. Nucl. Phys.} 67:140 (2012)

\bibitem{Kruger:2013kua}
Kr\"uger T, Tews I, Hebeler K, Schwenk A.
\newblock \textit{Phys. Rev. C} 88:025802 (2013)

\bibitem{Machleidt:2000ge}
Machleidt R.
\newblock \textit{Phys. Rev. C} 63:024001 (2001)

\bibitem{Wiringa:1994wb}
Wiringa RB, Stoks VGJ, Schiavilla R.
\newblock \textit{Phys. Rev. C} 51:38 (1995)

\bibitem{Wiringa:2002ja}
Wiringa RB, Pieper SC.
\newblock \textit{Phys. Rev. Lett.} 89:182501 (2002)

\bibitem{Carlson:1983kq}
Carlson J, Pandharipande VR, Wiringa RB.
\newblock \textit{Nucl. Phys. A} 401:59 (1983)

\bibitem{Pieper:2001ap}
Pieper SC, Pandharipande VR, Wiringa RB, Carlson J.
\newblock \textit{Phys. Rev. C} 64:014001 (2001)

\bibitem{Maris:2013rgq}
Maris P, et~al.
\newblock \textit{Phys. Rev. C} 87:054318 (2013)

\bibitem{Contessi:2017rww}
Contessi L, et~al.
\newblock \textit{Phys. Lett. B} 772:839 (2017)

\bibitem{Barnea:2013uqa}
Barnea N, et~al.
\newblock \textit{Phys. Rev. Lett.} 114:052501 (2015)

\bibitem{Bansal:2017pwn}
Bansal A, et~al.
\newblock \textit{Phys. Rev. C} 98:054301 (2018)

\bibitem{Weinberg:1990rz}
Weinberg S.
\newblock \textit{Phys. Lett. B} 251:288 (1990)

\bibitem{Weinberg:1991um}
Weinberg S.
\newblock \textit{Nucl. Phys. B} 363:3 (1991)

\bibitem{Kaplan:1998tg}
Kaplan DB, Savage MJ, Wise MB.
\newblock \textit{Phys. Lett. B} 424:390 (1998)

\bibitem{Kaplan:1998we}
Kaplan DB, Savage MJ, Wise MB.
\newblock \textit{Nucl. Phys. B} 534:329 (1998)

\bibitem{Nogga:2005hy}
Nogga A, Timmermans RGE, van Kolck U.
\newblock \textit{Phys. Rev. C} 72:054006 (2005)

\bibitem{PavonValderrama:2005wv}
Pavon~Valderrama M, Ruiz~Arriola E.
\newblock \textit{Phys. Rev. C} 74:054001 (2006)

\bibitem{Long:2011xw}
Long B, Yang CJ.
\newblock \textit{Phys. Rev. C} 85:034002 (2012)

\bibitem{vanKolck:1994yi}
van Kolck U.
\newblock \textit{Phys. Rev. C} 49:2932 (1994)

\bibitem{Reinert:2017usi}
Reinert P, Krebs H, Epelbaum E.
\newblock \textit{Eur. Phys. J. A} 54:86 (2018)

\bibitem{Entem:2017gor}
Entem DR, Machleidt R, Nosyk Y.
\newblock \textit{Phys. Rev. C} 96:024004 (2017)

\bibitem{Piarulli:2014bda}
Piarulli M, et~al.
\newblock \textit{Phys. Rev. C} 91:024003 (2015)

\bibitem{Ekstrom:2017koy}
Ekstr\"om A, et~al.
\newblock \textit{Phys. Rev. C} 97:024332 (2018)

\bibitem{Ekstrom:2015rta}
Ekstr\"{o}m A, et~al.
\newblock \textit{Phys. Rev. C} 91:051301 (2015)

\bibitem{Carlsson:2015vda}
Carlsson BD, et~al.
\newblock \textit{Phys. Rev. X} 6:011019 (2016)

\bibitem{Furnstahl:2015rha}
Furnstahl RJ, Klco N, Phillips DR, Wesolowski S.
\newblock \textit{Phys. Rev. C} 92:024005 (2015)

\bibitem{Melendez:2017phj}
Melendez JA, Wesolowski S, Furnstahl RJ.
\newblock \textit{Phys. Rev. C} 96:024003 (2017)

\bibitem{Gezerlis:2013ipa}
Gezerlis A, et~al.
\newblock \textit{Phys. Rev. Lett.} 111:032501 (2013)

\bibitem{Gezerlis:2014zia}
Gezerlis A, et~al.
\newblock \textit{Phys. Rev. C} 90:054323 (2014)

\bibitem{Piarulli:2016vel}
Piarulli M, et~al.
\newblock \textit{Phys. Rev. C} 94:054007 (2016)

\bibitem{Lovato:2011ij}
Lovato A, Benhar O, Fantoni S, Schmidt KE.
\newblock \textit{Phys. Rev.} C85:024003 (2012)

\bibitem{Huth:2017wzw}
Huth L, Tews I, Lynn JE, Schwenk A.
\newblock \textit{Phys. Rev. C} 96:054003 (2017)

\bibitem{Dyhdalo:2016ygz}
Dyhdalo A, Furnstahl RJ, Hebeler K, Tews I.
\newblock \textit{Phys. Rev. C} 94:034001 (2016)

\bibitem{Lynn:2015jua}
Lynn JE, et~al.
\newblock \textit{Phys. Rev. Lett.} 116:062501 (2016)

\bibitem{Epelbaum:2014efa}
Epelbaum E, Krebs H, Mei\ss{}ner UG.
\newblock \textit{Eur. Phys. J. A} 51:53 (2015)

\bibitem{Tews:2015ufa}
Tews I, Gandolfi S, Gezerlis A, Schwenk A.
\newblock \textit{Phys. Rev. C} 93:024305 (2016)

\bibitem{Shen:2012xz}
Shen G, et~al.
\newblock \textit{Phys. Rev. C} 86:035503 (2012)

\bibitem{Tanabashi:2018oca}
Tanabashi M, et~al.
\newblock \textit{Phys. Rev. D} 98:030001 (2018)

\bibitem{Carlson:1997qn}
Carlson J, Schiavilla R.
\newblock \textit{Rev. Mod. Phys.} 70:743 (1998)

\bibitem{Bernard:2001rs}
Bernard V, Elouadrhiri L, Mei\ss{}ner UG.
\newblock \textit{J. Phys. G} 28:R1 (2002)

\bibitem{Amaldi:1979vh}
Amaldi E, Fubini S, Furlan G.
\newblock \textit{Springer Tracts Mod. Phys.} 83:1 (1979)

\bibitem{Kitagaki:1983px}
Kitagaki T, et~al.
\newblock \textit{Phys. Rev. D} 28:436 (1983)

\bibitem{Meyer:2016oeg}
Meyer AS, Betancourt M, Gran R, Hill RJ.
\newblock \textit{Phys. Rev. D} 93:113015 (2016)

\bibitem{Rajan:2017lxk}
Gupta R, et~al.
\newblock \textit{Phys. Rev. D} 96:114503 (2017)

\bibitem{Marcucci:2001qs}
Marcucci LE, et~al.
\newblock \textit{Phys. Rev. C} 66:054003 (2002)

\bibitem{Kaiser:2003dr}
Kaiser N.
\newblock \textit{Phys. Rev. C} 67:027002 (2003)

\bibitem{Marcucci:2005zc}
Marcucci LE, et~al.
\newblock \textit{Phys. Rev. C} 72:014001 (2005)

\bibitem{Marcucci:2008mg}
Marcucci LE, et~al.
\newblock \textit{Phys. Rev. C} 78:065501 (2008)

\bibitem{Lovato:2013cua}
Lovato A, et~al.
\newblock \textit{Phys. Rev. Lett.} 111:092501 (2013)

\bibitem{Lovato:2015qka}
Lovato A, et~al.
\newblock \textit{Phys. Rev. C} 91:062501 (2015)

\bibitem{Lovato:2016gkq}
Lovato A, et~al.
\newblock \textit{Phys. Rev. Lett.} 117:082501 (2016)

\bibitem{Riska:1989bh}
Riska DO.
\newblock \textit{Phys. Rept.} 181:207 (1989)

\bibitem{Berg:1980lwp}
Berg H, et~al.
\newblock \textit{Nucl. Phys. A} 334:21 (1980)

\bibitem{Carlson:1991}
Carlson J, Pandharipande VR, Schiavilla R (1991).
\newblock Many-body Theory of Electron-nucleus Scattering: Light Nuclei,
  chap.~II.
\newblock World Scientific,  177--218

\bibitem{Bacca:2014tla}
Bacca S, Pastore S.
\newblock \textit{J. Phys. G} 41:123002 (2014)

\bibitem{Epelbaum:2002vt}
Epelbaum E, et~al.
\newblock \textit{Phys. Rev. C} 66:064001 (2002)

\bibitem{Gazit:2008ma}
Gazit D, Quaglioni S, Navratil P.
\newblock \textit{Phys. Rev. Lett.} 103:102502 (2009)

\bibitem{Walzl:2001vb}
Walzl M, Mei\ss{}ner UG.
\newblock \textit{Phys. Lett. B} 513:37 (2001)

\bibitem{Phillips:2003jz}
Phillips DR.
\newblock \textit{Phys. Lett. B} 567:12 (2003)

\bibitem{Pastore:2008ui}
Pastore S, Schiavilla R, Goity JL.
\newblock \textit{Phys. Rev. C} 78:064002 (2008)

\bibitem{Pastore:2009is}
Pastore S, et~al.
\newblock \textit{Phys. Rev. C} 80:034004 (2009)

\bibitem{Pastore:2011ip}
Pastore S, Girlanda L, Schiavilla R, Viviani M.
\newblock \textit{Phys. Rev. C} 84:024001 (2011)

\bibitem{Kolling:2009iq}
Kolling S, Epelbaum E, Krebs H, Meissner UG.
\newblock \textit{Phys. Rev.} C80:045502 (2009)

\bibitem{Kolling:2011mt}
Kolling S, Epelbaum E, Krebs H, Meissner UG.
\newblock \textit{Phys. Rev.} C84:054008 (2011)

\bibitem{Baroni:2015uza}
Baroni A, et~al.
\newblock \textit{Phys. Rev.} C93:015501 (2016), [Erratum: Phys.
  Rev.C95,no.5,059901(2017)]

\bibitem{Krebs:2016rqz}
Krebs H, Epelbaum E, Meißner UG.
\newblock \textit{Annals Phys.} 378:317 (2017)

\bibitem{Baroni:2016xll}
Baroni A, et~al.
\newblock \textit{Phys. Rev.} C94:024003 (2016), [Erratum: Phys.
  Rev.C95,no.5,059902(2017)]

\bibitem{Baroni:2017gtk}
Baroni A, Schiavilla R.
\newblock \textit{Phys. Rev.} C96:014002 (2017)

\bibitem{Baroni:2018fdn}
Baroni A, et~al.
\newblock \textit{Phys. Rev.} C98:044003 (2018)

\bibitem{Hammond:1994}
B.J.~Hammond W.A.~Lester PR.
\newblock World Scientific, Singapore (1994)

\bibitem{Nightingale:1999}
Nightingale M, Umrigar C.
\newblock Springer (1999)

\bibitem{Schmidt:1992}
Schmidt K, Ceperley D.
\newblock ed by K. Binder Springer, Berlin (1992)

\bibitem{Foulkes:2001}
Foulkes WMC, Mitas L, Needs RJ, Rajagopal G.
\newblock \textit{Rev. Mod. Phys.} 73:33 (2001)

\bibitem{Gandolfi:2009b}
Gandolfi S, et~al.
\newblock \textit{Phys. Rev. C} 80:045802 (2009)

\bibitem{Pudliner:1997}
Pudliner BS, et~al.
\newblock \textit{Phys. Rev. C} 56:1720 (1997)

\bibitem{Schmidt:1995}
Schmidt KE, Lee MA.
\newblock \textit{Phys. Rev. E} 51:5495 (1995)

\bibitem{Pieper:2004qw}
Pieper SC, Wiringa RB, Carlson J.
\newblock \textit{Phys. Rev. C} 70:054325 (2004)

\bibitem{Schmidt:1999}
Schmidt KE, Fantoni S.
\newblock \textit{Phys. Lett. B} 446:99 (1999)

\bibitem{Sarsa:2003}
Sarsa A, Fantoni S, Schmidt KE, Pederiva F.
\newblock \textit{Phys. Rev. C} 68:024308 (2003)

\bibitem{Lonardoni:2018prc}
Lonardoni D, et~al.
\newblock \textit{Phys. Rev. C} 97:044318 (2018)

\bibitem{Nollett:2006su}
Nollett KM, et~al.
\newblock \textit{Phys. Rev. Lett.} 99:022502 (2007)

\bibitem{Lonardoni:2018prl}
Lonardoni D, et~al.
\newblock \textit{Phys. Rev. Lett.} 120:122502 (2018)

\bibitem{Piarulli:2017dwd}
Piarulli M, et~al.
\newblock \textit{Phys. Rev. Lett.} 120:052503 (2017)

\bibitem{Lonardoni:2017egu}
Lonardoni D, Lovato A, Pieper SC, Wiringa RB.
\newblock \textit{Phys. Rev. C} 96:024326 (2017)

\bibitem{Chen:2016bde}
Chen JW, Detmold W, Lynn JE, Schwenk A.
\newblock \textit{Phys. Rev. Lett.} 119:262502 (2017)

\bibitem{Lonardoni:2018sqo}
Lonardoni D, Gandolfi S, Wang XB, Carlson J.
\newblock \textit{Phys. Rev. C} 98:014322 (2018)

\bibitem{Subedi:2008zz}
Subedi R, et~al.
\newblock \textit{Science} 320:1476 (2008)

\bibitem{Korover:2014dma}
Korover I, et~al.
\newblock \textit{Phys. Rev. Lett.} 113:022501 (2014)

\bibitem{Hen:2014nza}
Hen O, et~al.
\newblock \textit{Science} 346:614 (2014)

\bibitem{Sarsa:2003zu}
Sarsa A, Fantoni S, Schmidt KE, Pederiva F.
\newblock \textit{Phys. Rev.} C68:024308 (2003)

\bibitem{Gandolfi:2009fj}
Gandolfi S, et~al.
\newblock \textit{Phys. Rev.} C79:054005 (2009)

\bibitem{Tews:2018chv}
Tews I, Margueron J, Reddy S.
\newblock \textit{Phys. Rev.} C98:045804 (2018)

\bibitem{Gandolfi:2011xu}
Gandolfi S, Carlson J, Reddy S.
\newblock \textit{Phys. Rev. C} 85:032801 (2012)

\bibitem{Read:2008iy}
Read JS, Lackey BD, Owen BJ, Friedman JL.
\newblock \textit{Phys. Rev.} D79:124032 (2009)

\bibitem{Hebeler:2010jx}
Hebeler K, Lattimer JM, Pethick CJ, Schwenk A.
\newblock \textit{Phys. Rev. Lett.} 105:161102 (2010)

\bibitem{Alford:2015dpa}
Alford MG, et~al.
\newblock \textit{Phys. Rev.} D92:083002 (2015)

\bibitem{NICER}
Arzoumanian Z, Gendreau KC, Baker CL, et~al.
\newblock \textit{Proc. SPIE} 9144:9144 (2014)

\bibitem{Watts:2018iom}
Watts AL, et~al.
\newblock \textit{Sci. China Phys. Mech. Astron.} 62:29503 (2019)

\bibitem{Demorest:2010}
Demorest PB, et~al.
\newblock \textit{Nature} 467:1081 (2010)

\bibitem{Antoniadis13}
{Antoniadis et al.} J.
\newblock \textit{Science} 340:448 (2013)

\bibitem{TheLIGOScientific:2017qsa}
Abbott B, et~al.
\newblock \textit{Phys. Rev. Lett.} 119:161101 (2017)

\bibitem{Abbott:2018wiz}
Abbott BP, et~al.  arXiv:1805.11579 [gr-qc] (2018)

\bibitem{mb_web}
The MicroBooNE Experiment.
\newblock \url{http://www-microboone.fnal.gov}

\bibitem{nova_web}
The {NO}v{A} {E}xperiment.
\newblock \url{http://www-nova.fnal.gov}

\bibitem{t2k_web}
The {T}2{K} {E}xperiment.
\newblock \url{http://t2k-experiment.org}

\bibitem{dune_web}
The Deep Underground Neutrino Experiment.
\newblock \url{http://www.dunescience.org}

\bibitem{hk_web}
Hyper-Kamiokande.
\newblock \url{http://www.hyperk.org}

\bibitem{Benhar:2006er}
Benhar O, Day D, Sick I  arXiv:nucl-ex/0603032 [nucl-ex] (2006)

\bibitem{jlab_web}
Thomas Jefferson National Accelerator Facility.
\newblock \url{https://www.jlab.org}

\bibitem{Hen:2016kwk}
Hen O, Miller GA, Piasetzky E, Weinstein LB.
\newblock \textit{Rev. Mod. Phys.} 89:045002 (2017)

\bibitem{Rocco:2018mwt}
Rocco N, et~al.  arXiv:1810.07647 [nucl-th] (2018)

\bibitem{Lovato:2014eva}
Lovato A, et~al.
\newblock \textit{Phys. Rev. Lett.} 112:182502 (2014)

\bibitem{Benhar:2015ula}
Benhar O, Lovato A, Rocco N.
\newblock \textit{Phys. Rev.} C92:024602 (2015)

\bibitem{Piasetzky:2006ai}
Piasetzky E, et~al.
\newblock \textit{Phys. Rev. Lett.} 97:162504 (2006)

\bibitem{Schiavilla:2006xx}
Schiavilla R, Wiringa RB, Pieper SC, Carlson J.
\newblock \textit{Phys. Rev. Lett.} 98:132501 (2007)

\bibitem{Alvioli:2007zz}
Alvioli M, Ciofi~degli Atti C, Morita H.
\newblock \textit{Phys. Rev. Lett.} 100:162503 (2008)

\bibitem{Wiringa:2008dn}
Wiringa RB, Schiavilla R, Pieper SC, Carlson J.
\newblock \textit{Phys. Rev.} C78:021001 (2008)

\bibitem{Weiss:2018tbu}
Weiss R, et~al.  arXiv:1806.10217 [nucl-th] (2018)

\bibitem{Orlandini:2016hsk}
Orlandini G, Turro F.
\newblock \textit{Few Body Syst.} 58:76 (2017)

\bibitem{Carlson:1992ga}
Carlson J, Schiavilla R.
\newblock \textit{Phys. Rev. Lett.} 68:3682 (1992)

\bibitem{Carlson:2001mp}
Carlson J, Jourdan J, Schiavilla R, Sick I.
\newblock \textit{Phys. Rev.} C65:024002 (2002)

\bibitem{Lovato:2017cux}
Lovato A, et~al.
\newblock \textit{Phys. Rev.} C97:022502 (2018)

\bibitem{Carlson:1987}
Carlson J.
\newblock \textit{Phys. Rev. C} 36:2026 (1987)

\bibitem{Bryan1990}
Bryan RK.
\newblock \textit{European Biophysics Journal} 18:165 (1990)

\bibitem{Jarrell:1996rrw}
Jarrell M, Gubernatis JE.
\newblock \textit{Phys. Rept.} 269:133 (1996)

\bibitem{Jourdan:1996np}
Jourdan J.
\newblock \textit{Nucl. Phys.} A603:117 (1996)

\bibitem{Cloet:2015tha}
Cloët IC, Bentz W, Thomas AW.
\newblock \textit{Phys. Rev. Lett.} 116:032701 (2016)

\bibitem{Towner:1999}
Towner IS, Hardy JC :338 (1999), edited by P.\ Herczeg, C.M.\ Hoffman, and
  H.V.\ Klapdor-Kleingrothaus (World Scientific, Singapore)

\bibitem{Rocco:2018tes}
Rocco N, Leidemann W, Lovato A, Orlandini G.
\newblock \textit{Phys. Rev.} C97:055501 (2018)

\bibitem{Madeira:2018ykd}
Madeira L, Lovato A, Pederiva F, Schmidt KE.
\newblock \textit{Phys. Rev.} C98:034005 (2018)

\end{thebibliography}

\end{document}